\documentclass{article}

\usepackage[nonatbib,preprint]{neurips_2022}

\usepackage[utf8]{inputenc} 
\usepackage[T1]{fontenc}    
\usepackage{hyperref}       
\usepackage{url}            
\usepackage{booktabs}       
\usepackage{amsfonts}       
\usepackage{nicefrac}       
\usepackage{microtype}      
\usepackage{xcolor}         
\usepackage{tabto}    
\usepackage[superscript,biblabel]{cite}
\usepackage{setspace}  
\usepackage{titlesec}

\usepackage{subfig}
\usepackage{wrapfig}
\usepackage{tikz}
\usepackage{tikz-cd}
\usetikzlibrary{cd}
\usepackage{microtype}
\usepackage{graphicx}
\usepackage{booktabs} 
\usepackage{pifont}       
\usepackage{bbding}       %
\usepackage{fontawesome}  
\usepackage{jabbrv} 

\usepackage{amsmath}
\usepackage{amssymb}
\usepackage{mathtools}
\usepackage{amsthm}
\usepackage{arydshln}
\usepackage[capitalize,noabbrev]{cleveref}

\makeatletter
\renewcommand{\fnum@figure}{\textbf{Fig. \thefigure} | }
\renewcommand{\fnum@table}{\textbf{Table \thetable \ |}}

\makeatother
\usepackage[labelsep=space]{caption}
\usepackage[
singlelinecheck=false 
]{caption}

\usepackage{authblk}

\title{
Accurate transition state generation with an object-aware equivariant elementary reaction diffusion model
}

\author[1, 2, *]{Chenru Duan}
\author[3]{Yuanqi Du}
\author[1, 2]{Haojun Jia}
\author[1, 2]{Heather J. Kulik}
\affil[1]{Department of Chemistry, Massachusetts Institute of Technology, Cambridge, MA, 02139}
\affil[2]{Department of Chemical Engineering, Massachusetts Institute of Technology, Cambridge, MA, 02139}
\affil[3]{Department of Computer Science, Cornell University, Ithaca, NY, 14850}
\affil[*]{Corresponding to: duanchenru@gmail.com}

\begin{document}

\maketitle

\begin{abstract}
Transition state (TS) search is key in chemistry for elucidating reaction mechanisms and exploring reaction networks. 
The search for accurate 3D TS structures, however, requires numerous computationally intensive quantum chemistry calculations due to the complexity of potential energy surfaces. 
Here, we developed an object-aware SE(3) equivariant diffusion model that satisfies all physical symmetries and constraints for generating sets of structures – reactant, TS, and product – in an elementary reaction. 
Provided reactant and product, this model generates a TS structure in seconds instead of hours, which is typically required when performing quantum chemistry-based optimizations. 
The generated TS structures achieve a median of 0.08 Å root mean square deviation compared to the true TS. 
With a confidence scoring model for uncertainty quantification, we approach an accuracy required for reaction barrier estimation (2.6 kcal/mol) by only performing quantum chemistry-based optimizations on 14\% of the most challenging reactions. 
We envision the proposed approach useful in constructing large reaction networks with unknown mechanisms.
\end{abstract}

\setstretch{1.8}

\section*{Introduction}

Breaking down complex chemical reactions into their constituent elementary reactions is key for understanding reaction mechanisms and designing processes that favor target reaction pathways.\cite{ZimmermanWIREsRev,ReactNetworkAnnualRev,TSTReview} 
Due to the transient nature of the intermediate and transition state (TS) involved in these elementary reactions, it is difficult to isolate and characterize these structures experimentally.
Instead, high throughput quantum chemistry computation, for example, with density functional theory (DFT) \cite{DFTReview}, provides valuable insights on potential reaction mechanisms by constructing comprehensive reaction networks.\cite{ReactNetworkAnnualRev,Durant1996}
These networks are established by either iteratively enumerating potential elementary reactions on-the-fly given existing species\cite{Reiher2018,ReiherMRNetwork} or propagating biased \textit{ab initio} molecular dynamics followed by elementary reaction refinement\cite{Nanoreactor,NA-Nanoreactor,Zeng2020}.
Both approaches, however, require a tremendous number of quantum chemistry calculations due to the large number of species potentially involved in a chemical reaction.\cite{KinBot,vonLilienfeld2020, Margraf2023}

\qquad Among all DFT energy evaluations, the overwhelming majority comes from locating an accurate TS structure solely based on reactant and product information.\cite{ReactNetworkAnnualRev,TSTReview}
Nonetheless, obtaining these TS structures is vital for estimating reaction rates and determining dominant reaction pathways in a reaction network.
Conventional TS search algorithms (for example, nudged elastic band\cite{NEB}, or NEB) are computationally intensive and notorious for their difficulty in convergence\cite{ts1x}, yielding low success rates and wasting substantial computational resources\cite{Zhao2021}.
Recently, there has been growing interest in exploring the use of machine learning techniques for TS search.
This includes ideas that formulate TS search as a 2D graph-to-structure conversion problem \cite{G2S}, a "shooting game" solved by reinforcement learning \cite{RLTS}, generative tasks addressed alternately by graph neural networks (GNN) \cite{GreenPCCP}, a generative-adversarial network \cite{TSGAN} and a combination of gated recurrent neural network and transformer\cite{ChoiNatComm}, and using an ML potential as a surrogate for DFT during TS optimizations \cite{NeuralNEB}.
However, these approaches do not respect all the physical symmetries in describing an elementary reaction and require further reconstruction and optimization to obtain the final 3D TS structure.
In addition, they are still far from reaching the high precision required (that is, 3 kcal/mol, corresponding to a change of one order of magnitude in reaction rate at 300 °C) for estimating a TS barrier height in lieu of DFT evaluation\cite{TSTReview}.

\qquad Diffusion models \cite{ddpm,diffusion2015,scoresde} have recently been adapted in physical science problems, such as generating organic molecules \cite{EDM} and their conformations, protein-ligand docking \cite{DiffDock}, and structural-based drug design \cite{SBDD}.
There, an SE(3) equivariant GNN is used as the scoring function to preserve the required permutation, \textcolor{black}{translation}, and rotation symmetry for a 3D object (for example, molecule or protein) in the Euclidean space, which works ideally for systems that contain only one single object \cite{e3nn, EGNN, Nequip}.
However, there are many scenarios in chemistry and materials science where the desired system  consists of multiple objects, for which the relative positioning does not influence the system itself. 
This includes the design of compounds with multiple building blocks (for example, metal organic frameworks \cite{MOFChemRev}), pairs of molecules that have similar chemistry but demonstrate distinct properties (for example, the well-known activity cliff in protein binding \cite{ActCliff}), and chemical processes that involve multiple distinct structures such as in chemical reactions \cite{ReactNetworkAnnualRev}.
Existing diffusion models with SE(3) equivariant GNNs are problematic for modeling these systems as they do not respect all symmetries and constraints for describing these systems.

\qquad In this work, we developed a general procedure to adapt an SE(3) equivariant neural network to preserve all desired symmetries and constraints on systems that consist of multiple objects.
\textcolor{black}{
These symmetries include permutation among atoms in a fragment of reactant or product, permutation among fragments in reactant or product, and rotation and translation for each fragment in reactant and product (Supplementary Text \ref{SI:required_symmetries}).
With all these symmetries satisfied, we bypassed the need of atom order mapping and fragment alignment in TS search, generating TS structure with only the 3D geometry of fragments in reactant and product.
}
We demonstrated this "\underline{o}bject-\underline{a}ware" SE(3) GNN for generating sets of 3D molecules in elementary \underline{react}ions under the \underline{diff}usion model framework, which we refer to as OA-ReactDiff.
In particular, we focused on TS search, an essential but computationally demanding step for estimating reaction barrier heights, rates, and exploring reaction networks.
With OA-ReactDiff, the predicted TS structures are highly similar to the true TS structures with an average root mean square deviation (RMSD) of 0.18 Å within 6 seconds on a single GPU. 
We further built a recommender based on confidence ranking to select among samples generated by OA-ReactDiff, which reduced the average RMSD to 0.13 Å. 
Using the self-confidence score of OA-ReactDiff for uncertainty quantification, we obtain \textcolor{black}{a mean absolute error (MAE)} 2.6 kcal/mol on barrier height by only performing 14\% of the DFT-based optimizations for the most challenging systems for the model, approaching the accuracy required (that is, 3 kcal/mol) for exploring reaction networks with unknown mechanisms\cite{TSTReview}.
The high accuracy of generated 3D structures and reaction barrier estimate achieved by OA-ReactDiff provides the possibility of accelerating and even circumventing expensive quantum chemistry calculations normally required for TS search.

\section*{Results}

\begin{figure*}[t!]
    \centering
    \includegraphics[width=0.85\textwidth]{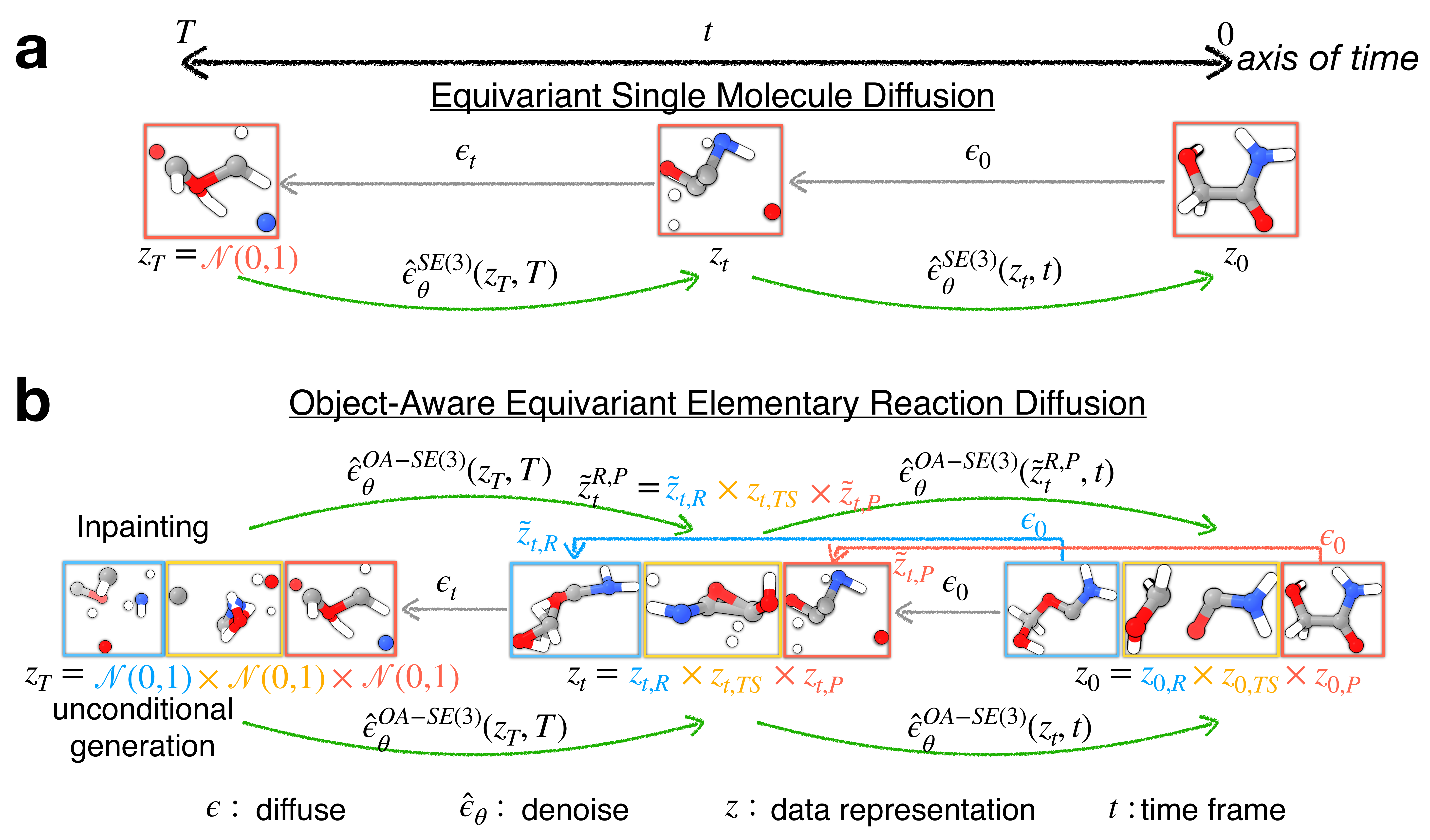}
    \caption{\textbf{Overview of equivariant diffusion models for generative molecular system sampling.}
    \textbf{a.} Equivariant diffusion model (EDM) for single molecule generation. In the diffusion process (gray arrows), the data distribution of molecules at time $t$ ($z_t$) diffuses by a Gaussian noise ($\epsilon_t$) from $t=0$ to $t=T$ until $z_T$ becomes a normal distribution. During the generation process (green arrows), a molecule is sampled from a normal distribution at $t=T$ after which an SE(3) equivariant denoising network ($\hat \epsilon_\theta^{SE(3)}$) is applied iteratively on the sample to remove noise, finally recovering the original data distribution at $t=0$. 
    \textbf{b.} Object-aware elementary reaction diffusion model (OA-ReactDiff), which generates a system as a joint distribution of multiple molecules that form a elementary reaction with reactant (R, blue), transition state structure (TS, orange), and product (P, red). Its diffusion process resembles that in EDM, while the sampling process requires that the denoising network respects object-level SE(3) equivariance (bottom). For conditional generation where part of the system is known \textit{a priori}, a combined scheme named inpainting, where diffusing on known parts (e.g, R and P) and denoising on unknown parts (for example, TS), will be used for recovering the original conditional joint distribution (top). Atoms are colored as follows: gray for C, blue for N, red for O, and white for H.
    }
    \label{fig:overview}
\end{figure*}

\begin{figure*}[t!]
    \centering
    \includegraphics[width=0.85\textwidth]{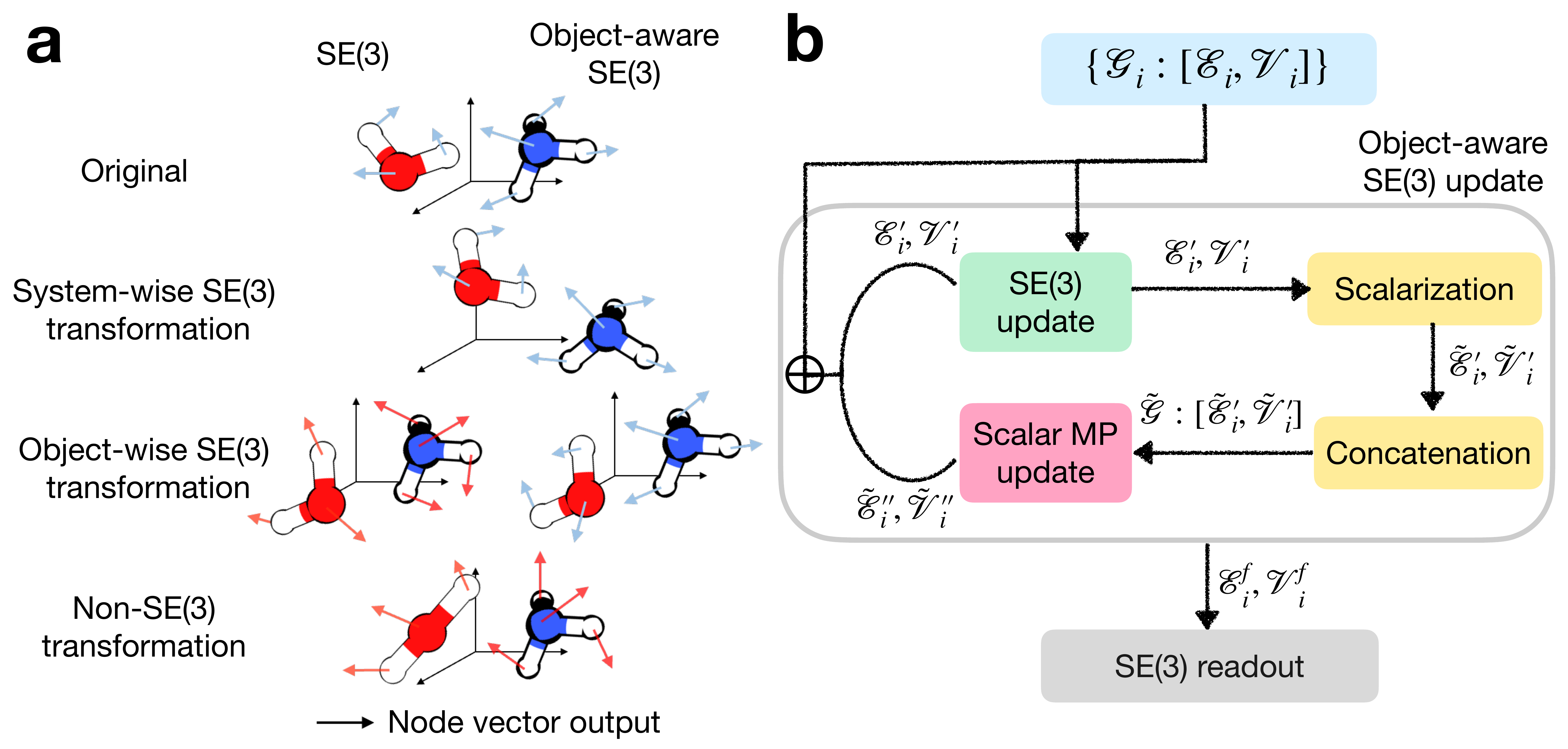}
    \caption{\textbf{Object-aware SE(3) equivariance and its implementation based on SE(3) equivariant graph neural networks.}
    \textbf{a.} Behavior of expected outputs for different transformations (suggested by different rows) on system with SE(3) (left) and object-aware SE(3) equivariance (right). Results that demonstrate the same behavior after transformation are grouped and only shown once for clarity. Node outputs that are preserved by symmetry (that is, are the same as the original data) are represented by blue arrows, otherwise by red arrows. Atoms are colored as follows: blue for N, red for O, and white for H.
    \textbf{b.} Construction of Object-aware SE(3) equivariant graph neural networks. A system consists of multiple objects ($i$) are represented by a set of subgraphs $\mathcal{G}_i$ with edges $\mathcal{E}_i$ and nodes $\mathcal{V}_i$, which can be both scalars (for example, atom types), vectors (for example, Cartesian coordinates), or higher-order tensors. This set of subgraphs first go through an SE(3) equivariant block for message passing and updating both their scalar and higher-order tensor features (green). The resulting high-order tensors ($\mathcal{E}_i^\prime$ and $\mathcal{V}_i^\prime$) of all objects are scalarized and concatenated as a system-level fully-connected graph with only scalar representation $\Tilde{\mathcal{G}}$ (yellow). This graph is then processed by a scalar message-passing (MP) block to include interactions among different objects ($i$) in the system (pink). The updated nodes and edges are combined with the outputs from equivariant update block as the input for the next object-aware SE(3) interaction block. The process repeats several times until the final object representations are readout (gray). 
}
    \label{fig:OASE3}
    
\end{figure*}

\paragraph{Overview of OA-ReactDiff.}
A diffusion model contains two processes \cite{diffusion2015, ddpm, scoresde}. 
In the forward (that is, diffusion) pass, Gaussian noise is continuously added to the original data distribution, which, over time, becomes an approximately normal distribution (Fig. \ref{fig:overview}a).  
In the reverse (that is, sampling) pass, a random sample is drawn from the normal distribution, after which a denoising neural network is iteratively applied to remove noise, recovering the original data distribution. 
This denoising network is trained to predict the noise added to the original data distribution (see \textit{\nameref{edm}}).
Since a 3D molecule or macromolecule fulfills permutational, translational, and rotational symmetry, the denoising graph neural network (GNN) used in chemistry application requires SE(3) equivariance (Fig. \ref{fig:OASE3}a). 
For a molecule represented by atom types (that is, scalars) and their Cartesian coordinates (that is, vectors), as one applies an SE(3) transformation (for example, rotation), the predicted noise on atom types should be the same while that on coordinates should undergo the same transformation.

\qquad Despite the success of SE(3) GNN-based equivariant diffusion models (EDM) in many chemistry applications, they inherently lack the symmetries required for systems containing multiple objects (for example, molecules) whose interactions are independent of their coordinates in the 3D Euclidean space.
An elementary reaction, as our system of interest, consists of three objects: reactant, TS, and product.
If an SE(3) transformation (that is, rotation) is applied on one of these object (for example, reactant), the description of this elementary reaction should stay invariant/equivariant.  \cite{ReactNetworkAnnualRev, Reiher2018}.
In addition, for a reactant or product that has multiple fragments, SE(3) transformations on individual fragments should also have no influence on the elementary reaction. 
A vanilla SE(3) GNN, however, would take these object-wise SE(3) transformations as if the entire system undergoes a non-SE(3) transformation and would yield non-equivariant results, breaking the symmetry required to predict the noise on atom types and Cartesian coordinates in EDMs (Fig. \ref{fig:OASE3}a).

\qquad There, we model an elementary reaction as a joint distribution of the 3D structures of the reactant, TS, and product (Fig. \ref{fig:overview}b). 
The diffusion process is essentially the same as the vanilla EDM, where independent Gaussian noise is added to reactant, TS, and product until they become independent normal distributions.
In the denoising process, however, an object-aware SE(3) equivariant GNN is used to preserve correct physical symmetries and constraints in an elementary reaction (Supplementary Text \ref{SI:required_symmetries}).
We consider two denoising schemes.
One is unconditional generation where reactant, TS, and product are all sampled from the normal distribution, which can be used to generate new elementary reactions from scratch (Fig. \ref{fig:overview}b).
In chemistry, however, many important applications are targeted for conditional generation, where some information of an elementary reaction of interest is known \textit{a priori}.
For example, in double-ended TS search, the 3D structure of both reactant and product is known, and the task is to find the unique corresponding TS structure.
For these conditional generation tasks, we applied the inpainting scheme, which models the joint distribution of the reactant, TS, and product where the unknown objects are inpainted during the inference time\cite{Repaint}. 
In TS search, specifically, it combines distributions from the diffused reactant \textcolor{black}{and} product (that is, known parts) and denoised TS structure (\textcolor{black}{that is}, unknown parts) at each step before proceeding to the following denoising step (Fig. \ref{fig:overview}b, see \textit{\nameref{inpainting}}).

\qquad In an elementary reaction, any non-SE(3) transformation on a single object (for example, reactant) should simultaneously influence all three objects, while any object-based SE(3) transformation on reactant, TS, and product should not change a reaction (Fig. \ref{fig:OASE3}a). 
While a vanilla SE(3) GNN fulfills the former requirement, it violates the latter symmetry as it considers all atoms in a system as belonging to the same molecule (Supplementary Table \ref{SI:ablation}). 
Here, we achieve all required physical symmetries in elementary reactions by developing a general procedure to adapt any SE(3) equivariant GNN as object-aware SE(3) equivariant with minimal effort (Fig. \ref{fig:OASE3}b and Supplementary Text \ref{SI:required_symmetries}). 
In this procedure, we build an object-aware SE(3) interaction layer from a regular SE(3) update layer, a series of non-parameterized operations (i.e, scalarization and concatenation), and a scalar-only message-passing update. 
In essence, the SE(3) update only operates on individual objects, where the relative positioning is not encoded, to learn a comprehensive representation for each molecule, while the scalar-only message passing layer learns interactions among atoms from different molecules (see \textit{\nameref{oa_details}}).
Similar to standard SE(3) GNNs, this object-aware SE(3) update repeats several times until a final SE(3) readout layer to pool out the final predicted noise on atom types and Cartesian coordinates.
In this work, we choose LEFTNet\cite{leftnet}, our recently-developed SE(3) GNN that reaches comparable state-of-the-art performance on QM9 \cite{QM9} and MD17 \cite{MD17}, as the vanilla SE(3) GNN for OA-ReactDiff (see \textit{\nameref{leftnet}}).

\vskip 0.2in
\paragraph{OA-ReactDiff training.}
We trained OA-ReactDiff on Transition1x, \cite{ts1x} a dataset that contains climbing-image NEB \cite{CINEB} calculated reactant, TS structure, and product at the $\omega$B97x/6-31G(d) level of theory \cite{wb97x,631gs} on 10,073 organic reactions of various types originated from a quite exhaustive enumeration \cite{Grambow2020, GreenDataApp} of product-reactant pairs based on the GDB7 \cite{gdb17} dataset. 
Each reaction consists of up to seven heavy atoms including C, N, and O, with the largest system consisting of 23 total atoms.
The use of climbing-image NEB ensures a relatively accurate TS structure, making each elementary reaction in Transition1x a unique set of reactant, TS, and product, which guarantees the necessary condition for training OA-ReactDiff.
We trained OA-ReactDiff on 9,000 elementary reactions randomly partitioned from Transition1x, leaving 1,073 unseen reactions as the test set.
Despite the potential overlap of certain chemical species in the training and test set, there are always at least two species (reactant and TS or TS and product) that are distinct in any test reaction compared to all training data (see \textit{\nameref{training}}).

\qquad In OA-ReactDiff, a molecule is represented by atom types with one-hot encoding and nuclear charges and Cartesian coordinates of its constituent atoms.
It is common to consider all components of the atom representation in the diffusion and denoising process as none of them is considered known \textit{a priori}\cite{EDM, SBDD}.
In chemical reactions, however, it is reasonable to assume that we know the atom types due to the conservation of atoms \cite{DiffDock}.
Therefore, we only diffused and denoised the Cartesian coordinates of the reactant, TS structure, and product in OA-ReactDiff (Fig. \ref{fig:overview}b).
Since OA-ReactDiff satisfies all the symmetries and constraints for describing an elementary reaction, it does not require any pre-processing of reaction data, such as atom order matching for different species and careful alignment of reactants and products, which sometimes can be infeasible to obtain \cite{ChoiNatComm} (see \textit{\nameref{edm}} and \textit{\nameref{oa_details}}). 
\textcolor{black}{
Transition1x dataset has atom mapping and aligned product molecules as constructed.
To showcase of the capability of OA-ReactDiff not replying on the atom mapping and fragment alignment, we intentionally swapped the atom ordering and broke the alignment in Transition1x beforehand and verified that OA-ReactDiff functions well without these pre-processing requirement (Supplementary Figure \ref{Supp:example_multi_molecule_rxn}). 
}
Due to the use of an object-aware SE(3) GNN, OA-ReactDiff breaks the reflection symmetry and thus can distinguish chiral molecules (see \textit{\nameref{leftnet}}). 
OA-ReactDiff also bypasses the need for data augmentation in the case of reversing reaction direction by enforcing the same graph embedding layer for reactant and product, and thus guarantees the outputs are invariant to the order of reactant and product as inputs\cite{ChoiNatComm}. 
Lastly, there are no post-processing steps (for example, reconstructing the 3D structure from a distance matrix through optimizations) required as OA-ReactDiff directly yields the Cartesian coordinates of reactant, TS structure, and product \cite{G2S}.
These outstanding features make OA-ReactDiff an end-to-end model for elementary reaction generation and TS search.

\begin{figure*}[t!]
    \centering
    \includegraphics[width=1.00\textwidth]{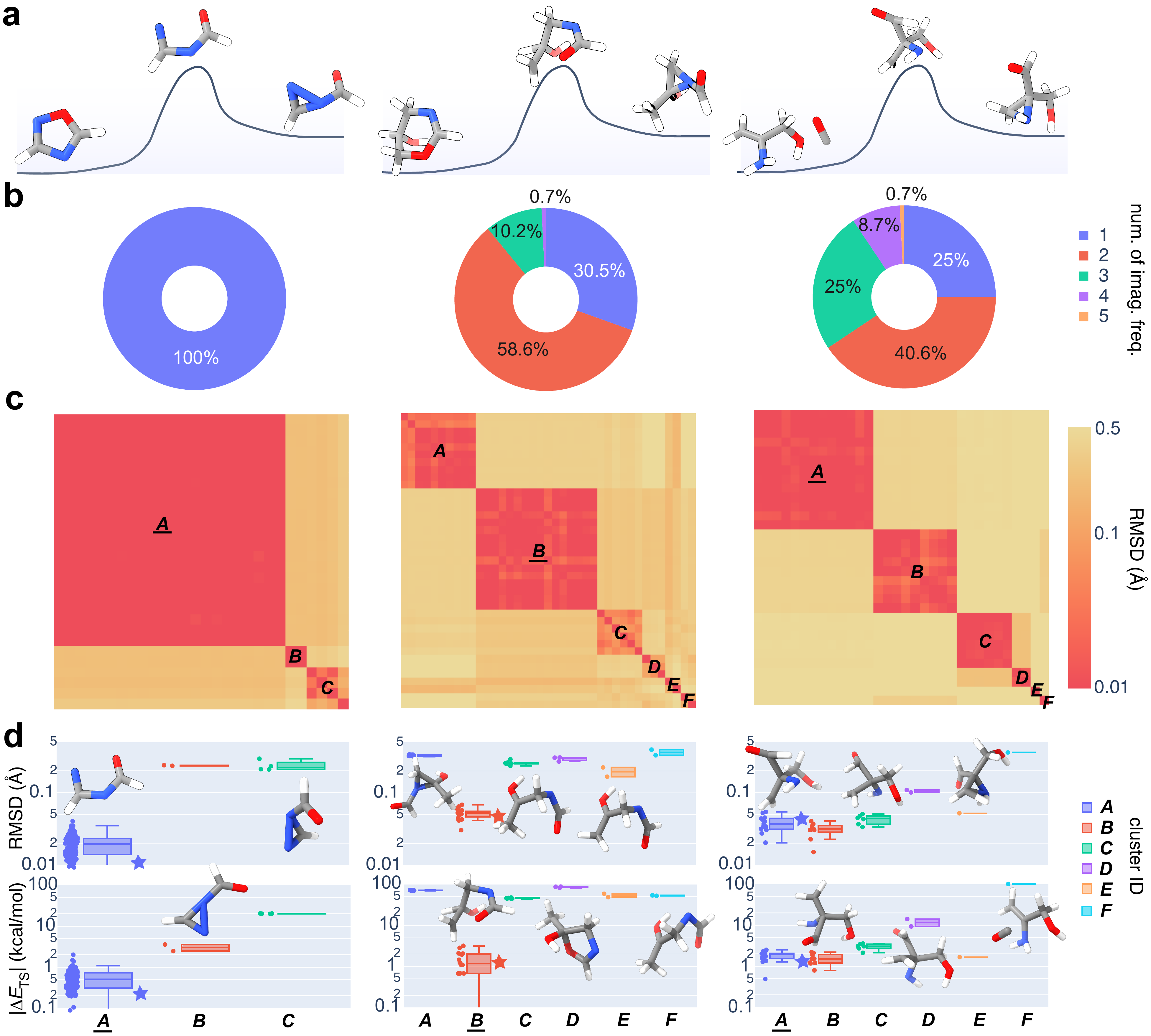}
    \caption{\textbf{Analysis of samples generated by OA-ReactDiff on select elementary reactions.}
    \textbf{a.} Illustration of select reactions: aromatic 5-membered ring to high-valence 3-membered ring rearrangement that breaks and forms one bond (left, b1f1), 6-membered ring to saturated 3-membered ring rearrangement (middle, b2f2), and carbonyl attack reaction (right, b2f3). All three elementary reactions have not been seen during the model training.
    \textcolor{black}{
        \textbf{b.} Pie chart showing the percentage of OA-ReactDiff generated TS structure with different numbers of imaginary frequencies (imag. freq.) evaluated by a Hessian calculation by $\omega$B97x/6-31G(d).
        \textbf{c.} Pairwise RMSD matrix for OA-ReactDiff generated and $\omega$B97x/6-31G(d)-optimized TS structures. Only the generated structures with one imag. freq. are optimized by DFT. The structures are sorted to form clusters (where the RMSDs are smaller than 0.1 Å) on the diagonal line of the matrix. Each cluster is labeled by a capital bold italic letter. The cluster corresponding to the true TS is underlined. RMSD is colored through a red-to-yellow color bar in the log scale to show the clustering effect more clearly.
        \textbf{d.} RMSD (top) and absolute energy difference ($|\Delta E_{\mathrm{TS}}|$, bottom) between OA-ReactDiff generated and their corresponding $\omega$B97x/6-31G(d)-optimized TS structures for each reaction. These structures are grouped by the cluster observed in \textbf{c.}. A representative structure for each cluster is shown. The RMSD and $|\Delta E_{\mathrm{TS}}|$ for top-1 confidence TS structure picked by the recommender are shown with the star symbol.
    }
    Atoms are colored as follows: gray for C, blue for N, red for O, and white for H.
    }
    \label{fig:example_generation}
    
\end{figure*}

\vskip 0.2in
\paragraph{Overcoming the stochastic nature of diffusion models with confidence ranking.}
OA-ReactDiff models the joint distribution of a set of reactant, TS, and product, and thus can generate new elementary reactions without any conditions, including for those which the chemical composition is unseen during the model training (Supplementary Figure \ref{Supp:unconditional_samples}). 
Evaluating the accuracy and building a reaction network from these generated elementary reactions, however, require substantial computational resources for running DFT optimizations and may be subject to selection bias on which chemical compositions are included during evaluation.
Therefore, we focus on evaluating OA-ReactDiff under the scheme of conditional generation, specifically for TS search where the task is to identify the 3D TS structure provided a pair of reactant and product.

\qquad We first consider three example reactions in Transition1x that break and form a varied number of bonds, representing different levels of complexity (Fig. \ref{fig:example_generation}a).
Due to the stochastic nature of diffusion models, sampled TS structures from OA-ReactDiff will not be unique with a fixed reactant-product pair.
For each of the three reactions, we ran the OA-ReactDiff under the inpainting scheme 128 times, generating 128 distinct samples.
\textcolor{black}{
We then computed the Hessian for these 128 generated TS structures with $\omega$B97x/6-31G(d) and found many only contained one imaginary frequency and thus are good candidates for single-ended TS optimizations (Fig. \ref{fig:example_generation}b).
}
To evaluate the differences among the \textcolor{black}{
structures with one imaginary frequency, we performed TS optimization on these structures with $\omega$B97x/6-31G(d).
We then computed the pairwise RMSD and identified multiple distinct TS structures discovered by OA-ReactDiff. (Fig. \ref{fig:example_generation}c).
Further internal reaction coordinate calculations with $\omega$B97x/6-31G(d) confirmed that these TS structures lead to different reactant and product conformation or connectivity, and thus form distinct elementary reactions compared to the intended one (Supplementary Data).
These generated TS structures sometimes show relatively large structure deviations (that is, > 0.2 Å) compared to the optimized TS structures, which is expected since the input reactant and product do not actually correspond to the generated TS structure (Fig. \ref{fig:example_generation}d).
However, these different TS structures can still be exploited during the reaction network exploration for elementary reactions that would otherwise be neglected.
}

\qquad \textcolor{black}{
Provided the 3D conformations of reactant and product, there is only one unique TS structure \cite{QiyuanNCS2021}.
Yet the stochastic nature of OA-ReactDiff generates TS structures in a non-deterministic manner.
}
To address this challenge, we further trained an object-aware SE(3) LEFTNet as a confidence model, \cite{AF2, DiffDock} which also satisfies all the symmetries and constraints for elementary reactions.
There, provided a set of input reactant, TS, and product, the confidence model predicts its probability of being a true elementary reaction.
During its training, we provided elementary reactions sampled by OA-ReactDiff, which are labeled as good (i.e, 1) if the RMSD between sampled and true TS structure is < 0.2 Å and bad (that is, 0) otherwise (see \textit{\nameref{training}}).
Once trained, the confidence model successfully distinguishes different 3D TS structures generated by OA-ReactDiff, assigning them a distinct probability score (Fig. \ref{fig:example_generation}d).
Moreover, the confidence model always give the highest probability score to generated structures with among the lowest RMSD with respect to the true TS structure for all three example reactions.
Without the confidence model, random selection from samples generated by OA-ReactDiff may \textcolor{black}{
result in a TS structure of completely different conformation or connectivity compared to the reactant and product.
It would in turn}
yield a large (> 10 kcal/mol) energy difference for the predicted and true TS structure, which would lead to orders of magnitude differences in predicted reaction rates.
Even though it is not guaranteed that the confidence model always selects the sample with the lowest RMSD compared to the true TS structure, the confidence model will likely avoid choosing samples that have incorrect connectivity or geometries with large deviations, especially in reactions that have multiple bonds breaking and forming.

\begin{figure*}[t!]
    \centering
    \includegraphics[width=0.85\textwidth]{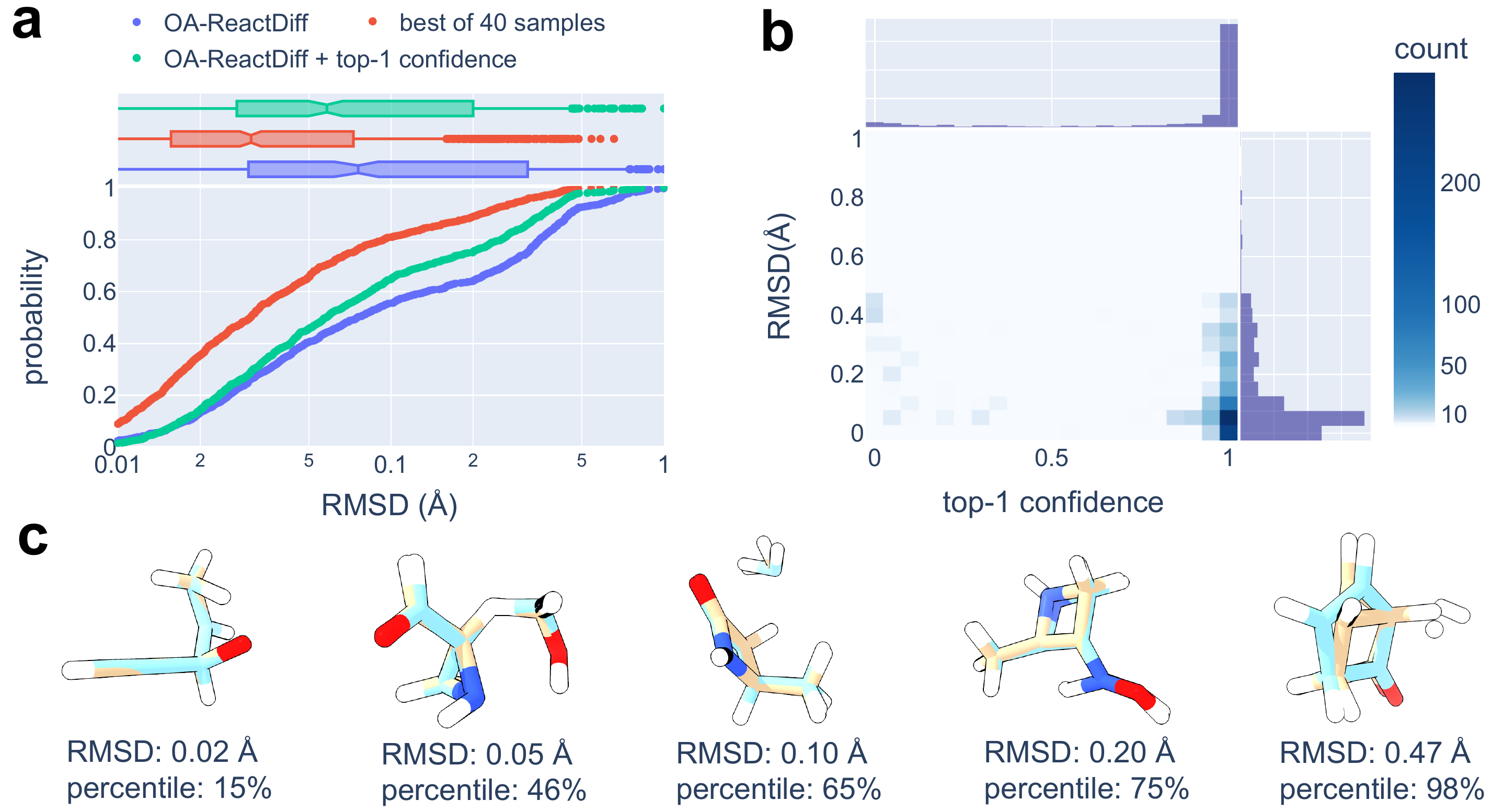}
    \caption{\textbf{Evaluation of structural similarities for TS structures generated by OA-ReactDiff and true TS structures.}
    \textbf{a.} Cumulative probability for RMSD between the true TS structures and OA-ReactDiff samples on 1073 set-aside test reactions. The OA-ReactDiff samples are evaluated under one-shot generation (blue), the top-1 confidence sample (that is, recommended, green), and the best (yet unachievable) sample out of 40 generated samples for each reaction (red). A log scale of the RMSD is presented for better visibility of the low-RMSD region.
    \textbf{b.} 2D density map for the RMSD vs. top-1 confidence for OA-ReactDiff generated samples. A log-scale color gradient is applied to the color bar to reveal low-density areas, which would otherwise be difficult to distinguish. 
    \textbf{c.} Example overlapping true and OA-ReactDiff + recommender TS structures at different RMSD and RMSD percentile rank shown in ascending order from left to right. Atoms are colored as follows: C in the true TS structure are in tan and those in the top-1 confident OA-ReactDiff sample are in skyblue; N for blue, O for red, and H for white.
    }
    \label{fig:RMSD}
    
\end{figure*}

\vskip 0.2in
\paragraph{High quality TS structures from OA-ReactDiff.}
We next systematically evaluated the structural similarity between the OA-ReactDiff and true TS structures for 1,073 set-aside unseen reactions in Transition1x, as judged by RMSD.
Notably, in contrast to an average runtime of 12 hours using climbing image NEB\cite{NeuralNEB} with DFT, it only takes \textcolor{black}{on average within 6 seconds} to generate a TS structure with OA-ReactDiff \textcolor{black}{
with proper batching on a V100 GPU (single sample generation without batching takes 17 seconds, Supplementary Table \ref{SI:runtime_vs_bs})
}.
Compared to bond lengths, angles, and dihedrals that mostly compare local geometry for a subset of atoms, the RMSD should provide a more accurate assessment on overall structural agreement, which is the ultimate goal of TS search\cite{ReactNetworkAnnualRev,ZimmermanWIREsRev}. 
For each reaction, we ran OA-ReactDiff 40 times, generating 40 independent guess TS structures. 
For a random selection of 40 samples, OA-ReactDiff has already reached an average RMSD of 0.183 Å with a median being 0.076 Å for the 1073 test elementary reactions (Fig. \ref{fig:RMSD}a).
More than half (two thirds) of the TS structures have an RMSD < 0.1 (0.2) Å compared to their corresponding true TS structures identified by climbing image NEB.
\textcolor{black}{
We observed a near power-law dependence of RMSD for OA-ReactDiff generated samples with the number of training data (Supplementary Table \ref{SI:training_data_scaling} and Fig. \ref{Supp:rmsd_scaling}).
The performance of OA-ReactDiff shows no sign of depletion when we reach the maximum of our training data available, suggesting room for further improvement through training OA-ReactDiff on larger datasets.
More interestingly, we find that, given a similar amount of training data, an OA-ReactDiff model trained only on system within 15 atoms performs similarly well to that trained on randomly sampled reactions (Supplementary Table \ref{SI:training_data_scaling}).
Meanwhile, the OA-ReactDiff model yields similar RMSD distribution regardless of system size (Supplementary Figure \ref{Supp:rmsd_size_box}).
This observation showcases great size extensiveness of OA-ReactDiff, which is encouraging for its application in practical reaction exploration for large molecules.
}

\qquad With the confidence model, we can further improve the procedure of sample selection using a recommender approach. \cite{DFARec}
Together with the true reactant and product, these guess TS structures are fed into the confidence model to get their probability score.
The sample with highest probability score (that is, top-1 confidence) is chosen as the final predicted TS structure from OA-ReactDiff.
With this recommender approach, the quality of selected TS structures is greatly improved, most likely due to the removal of TS structures with incorrect connectivity and geometries with large deviations (Fig. \ref{fig:example_generation}).
Moreover, the recommended structures mostly reside in the low RMSD and high confidence region, which demonstrates the effectiveness of our combined OA-ReactDiff and confidence recommender approach (Fig. \ref{fig:RMSD}b).
The average and median of the error in RMSD become 0.129 and 0.058 Å, respectively, with approximately two thirds of the recommended TS structures having RMSD < 0.1 Å (Fig \ref{fig:RMSD} a).
We observed a systematically-improving performance for the OA-ReactDiff + recommender approach as the number of total independent runs increases (Supplementary Figure \ref{Supp:metric_vs_runs}).
Here, we took 40 runs for each sample for a balance between total run time (4 minutes in total) and sampling accuracy.
Despite the fact that the recommender is still far from perfect for distinguishing structures with low RMSD (i.e, < 0.2 Å) structures, it helps avoiding TS samples that are very different from the true TS (that is, > 0.45 Å, Fig. \ref{fig:RMSD}c).
\textcolor{black}{
We also trained a regressor that predicts the RMSD between generated and true TS given a set of reactant, product, and generated TS structure as the confidence model, where the same quantitative behavior is observed (Supplementary Figure \ref{Supp:regressor_confidence}).
}
This recommender feature is particularly useful in end-to-end applications for ML models.

\begin{figure*}[t!]
    \centering
    \includegraphics[width=0.98\textwidth]{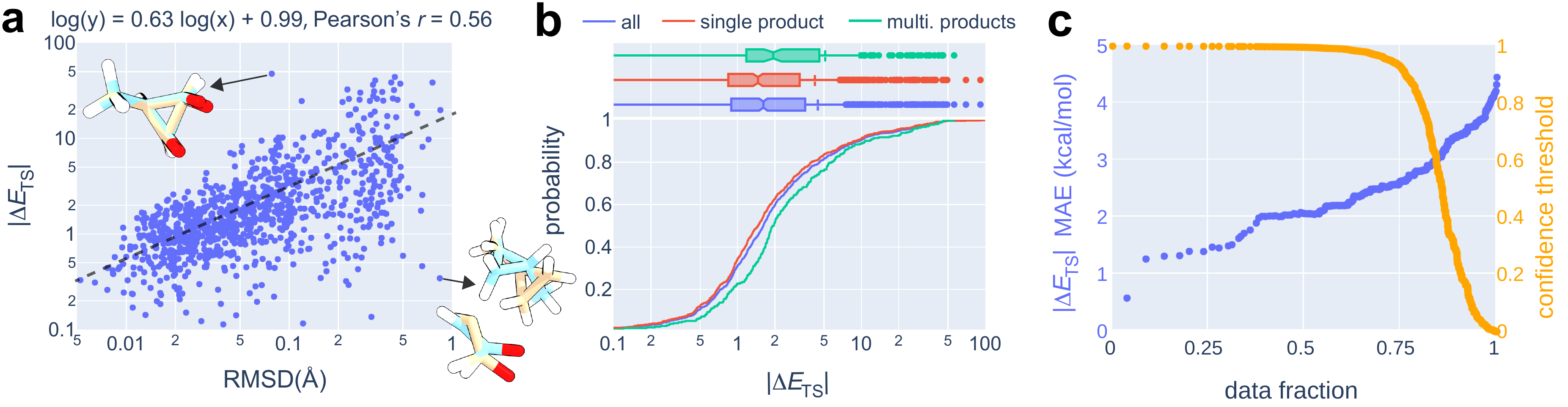}
    \caption{\textbf{Energetic performance for OA-ReactDiff + recommender TS structures.}
    \textbf{a.} Absolute energy error ($|\Delta \mathit{E}_\mathrm{TS}|$) vs. RMSD for top-1 confidence sample in log-log scale. A linear regressed black dashed line is shown with the fitted parameters at the top. Two example TS structures that deviate the most from this linear fit are shown.
    \textbf{b.} Cumulative probability for absolute energy error of top-1 confidence sample. The errors are evaluated by 1) all 1073 reactions in the test set (blue), 2) 783 reactions that involves only one single product (red), and 3) 290 reactions that yield multiple products (green). 
    \textbf{c.} MAE of $|\Delta \mathit{E}_\mathrm{TS}|$ (blue, left y-axis) and the corresponding confidence threshold (orange, right y-axis) as a function of the fraction of data considered in the 1073 test reactions.
    Atoms are colored as follows: C in the true TS structure are in tan and those in the top-1 confident OA-ReactDiff sample are in skyblue; N for blue, O for red, and H for white.
    }
    \label{fig:Ediff}
    
\end{figure*}

\vskip 0.2in
\paragraph{Approaching the energetic accuracy needed in TS search.}
Besides the exact 3D structure for a TS that provides insights into how an elementary reaction happens microscopically, a TS search algorithm should also evaluate the reaction barrier height, which is crucial for pruning large reaction networks and estimating reaction rates\cite{ReactNetworkAnnualRev}.
Here, we evaluate the performance of OA-ReactDiff on predicting the barrier height at the  DFT level of theory.
\textcolor{black}{
During the DFT evaluation, we used $\omega$B97x/6-31G(d) throughout this work, consistent with the method and basis set choice with Transition1x.
}
Specifically, we compared the electronic energy between OA-ReactDiff recommended and true TS structures \textcolor{black}{
for evaluating the barrier height.
}.
As one may expect, the absolute energy difference ($|\Delta \mathit{E}_\mathrm{TS}|$) between OA-ReactDiff and true TS structures has positive correlation with their RMSD.
We find they follow a power law best across various type of common algebraic fits, giving a Pearson's \textit{r} of 0.56 on a log-log plot (Fig. \ref{fig:Ediff}a).
This relatively low Pearson's \textit{r} can be explained by the high complexity of potential energy surfaces for molecules, where the direction of displacing an atom has large influences on the energy change.
For example, an OA-ReactDiff TS of $\mathrm{C_4 H_6 O_2}$ that has a relatively low RMSD of 0.078 Å compared to the true TS was found to have the highest energy difference (49.3 kcal/mol) among the 1073 test reactions(Fig. \ref{fig:Ediff}a).
This counter-intuitive result, however, is an artifact due to self-consistent field calculations being converged to different local minima in DFT energy evaluation for OA-ReactDiff and true TS (Supplementary Figure \ref{Supp:C2H6O2}).
On the other hand, despite an extremely large RMSD (0.821 Å) between OA-ReactDiff and the true TS structure for $\mathrm{C_6 H_{10} O}$, the energy difference between the two structures is only 0.4 kcal/mol, due to the fact that this TS consists of two fragments that only weakly interact with each other (Fig. \ref{fig:Ediff}a). 
There, OA-ReactDiff provides quite precise geometry for both fragments, in spite of the incorrect orientation between the two fragments, leading to a small deviation in barrier estimation (Supplementary Figure \ref{Supp:C6H10O}).

\qquad With OA-ReactDiff and the recommender, we reach an average of 4.4 kcal/mol and median of 1.6 kcal/mol for the absolute energy difference between the generated and true TS structure, with 71\% of TS barrier errors < 3 kcal/mol  (Fig. \ref{fig:Ediff}b).
OA-ReactDiff + recommender far outperforms semi-empirical methods such as density functional tight binding\cite{DFTB}, which \textcolor{black}{
was reported to have an MAE of 16.1 kcal/mol and average runtime of 82 seconds on Transition1x dataset.\cite{NeuralNEB}
}
Interestingly, the MAE would only improve marginally to 4.0 kcal/mol if we were able to select the OA-ReactDiff sample with the lowest RMSD compared to the true TS, indicating the power of the recommender for selecting TS structures with low energy deviations (Table \ref{table:EvsRMSD}).
The performance of OA-ReactDiff with the recommender on elementary reactions with multiple reactants and/or products is comparable to the performance for rearrangement reactions that only contain one single reactant and product (Supplementary Figure \ref{Supp:num_prod}) .
The slight deterioration of the performance is likely due to the imbalance of reaction types included in Transition1x, where only one fourth of the elementary reactions contain multiple reactants or products.
\textcolor{black}{
It is known that the TS structure is less sensitive to the choice of DFT functional\cite{Goddard2012}, especially or small organic molecules with only CNOH.
Here, we also evaluated the error for barrier height estimation of OA-ReactDiff with two other functionals and found quantitatively similar performance (Supplementary Table \ref{SI:method_dependence}).
}

\qquad We also compare OA-ReactDiff + recommender with two pioneering works where non-diffusion-based approaches were developed for generating 3D TS structures on large diverse organic reaction datasets such as Transition1x \cite{ts1x} or its predecessor \cite{Grambow2020}. 
Choi developed a "PSI-based" model\cite{ChoiNatComm} combining transformer and bidirectional gated recurrent unit that generates TS structures from refining the linear interpolation of reactant and product, which, however, requires the prior knowledge of atom mapping and \textcolor{black}{careful alignment among fragments in reactant and product for multi-molecular reactions.}
Schreiner \textit{et al.} \cite{NeuralNEB} trained a machine learning potential on 10M structures (with both energy and forces) collected during the generation of Transition1x, and, for the first time, applied the trained potential to the TS search problem in place of DFT.
There, similar to the problem of DFT-based TS search (for example, NEB), an attempt may still encounter convergence issues during the saddle point optimization, leading to a null prediction for the final TS structure.
We find OA-ReactDiff + recommender systematically outperforms the prior approaches on both the RMSD and barrier height estimate in terms of both the mean and median of the error distribution (Table \ref{table:EvsRMSD}).
This superior performance is attributed the fact that OA-ReactDiff manages to respect all physical symmetries and constraints for describing an elementary reaction, without the need for atom order mapping, reactants or products alignment, reconstruction of 3D geometry from distance matrix, and data augmentation of any kind.
\textcolor{black}{
We ascribe the slightly lower Pearson’s \textit{r} between RMSD and barrier height error of OA-ReactDiff to the presence of many high RMSD but low barrier height TS samples, which correspond to cases where OA-ReactDiff can generate good 3D structures for nearly non-interacting individual fragments but not their alignment (Fig. \ref{fig:Ediff}a and Supplementary Figure \ref{Supp:fit_compare}).
}
In addition, a more gradual increase of absolute energy difference with respect to RMSD was identified in OA-ReactDiff + recommender compared with the two other approaches, suggesting a more accurate barrier estimate can be obtained by OA-ReactDiff at the same level of structural similarity between generated and true TS (Table \ref{table:EvsRMSD} and Supplementary Figure \ref{Supp:fit_compare}).

\qquad We would ideally aim to select one single TS structure sampled by OA-ReactDiff with the recommender.
The recommended sample, however, may not be confident if all 40 samples generated by OA-ReactDiff suffer from a low confidence score (i.e, $p$ < 0.5) due to the limited amount of training data.
Further removal of these reactions (153, or 14\%) from the test set leads to a substantially improved energy difference with a mean of 3.1 kcal/mol and median of 1.4 kcal/mol (Table \ref{table:EvsRMSD}).
Moreover, we observe a monotonic behavior between the MAE for barrier height estimates and the confidence threshold imposed for TS structure generation that we consider as valid (Fig. \ref{fig:Ediff}c).
This desired monotonic behavior suggests that we can use the confidence score for uncertainty quantification to balance the accuracy and number of DFT calculations required in a practical workflow \cite{DuanUQWorkflow} that combines OA-ReactDiff, recommender, and DFT-based TS search.
For a set of TS structures generated by OA-ReactDiff and their corresponding confidence score evaluated by the confidence model and recommender, we can decide whether we would accept the recommended TS structure depending on its confidence score or would rather launch a DFT-based NEB.
With a confidence threshold of 0.5, we would only perform NEB with DFT on 14\% of reactions while directly accepting TS structures from OA-ReactDiff + recommender for the remaining 86\% reactions, leading to an overall accuracy of 2.6 kcal/mol.
This strategy showcases the power of combining OA-ReactDiff, recommender, and DFT-based NEB for efficient generation of TS structures given a target accuracy level.

\begin{table*}[th]
\centering 
\caption{\textbf{Summary of statistics for RMSD and absolute energy differences of TS structures and the linear fitting results obtained by various approaches.} 
The linear fit is obtained between $\mathrm{log}(|\Delta E_{\mathrm{TS}}|)$ vs. $\mathrm{log(RMSD)}$.
Some approaches invoke uncertainty quantification (OA-ReactDiff with $p$ > 0.5) or subject to convergence issues during optimization (PSI-based model and NeuralNEB), leading to inconclusive predictions some elementary reactions and thus resulting an effective data fraction < 1. 
\textcolor{black}{
Note that PSI-based model\cite{ChoiNatComm} used dataset from Grambow et al.\cite{Grambow2020}, which is the predecessor of Transition1x. 
Although all three approaches used roughly 8/2 as training/test ratio, the random seeds for the partition are not the same, which may have influence on the comparison.
}
}
\resizebox{1.00\textwidth}{!}{
\begin{tabular}{l|cc|cc|ccc|c}\toprule
\multicolumn{1}{c|}{Approach}
&\multicolumn{2}{c|}{RMSD (Å)} &\multicolumn{2}{c|}{$|\Delta E_{\mathrm{TS}}|$ (kcal/mol)} &\multicolumn{3}{c|}{fitting coefficeint}&\multicolumn{1}{c}{data fraction}\\\midrule
&mean &median &mean &median &slope &intercept&Pearson's \textit{r}\\\midrule
OA-ReactDiff&0.183&0.076&6.2&1.7&0.65&0.99&0.59&1.00\\
\hline  \hline
OA-ReactDiff + rec.&0.129&0.058&4.4&1.6&0.63&0.99&0.56&1.00\\
OA-ReactDiff + rec. 
($p$ > 0.5)&0.106&0.047&3.1&1.4&0.55&0.88&0.53&0.86\\
\hline \hline
OA-ReactDiff + best&0.071&0.031&4.0&1.3&0.72&1.19&0.65&1.00\\
OA-ReactDiff + best ($p$ > 0.5)&0.044&0.021&1.8&1.1&0.56&0.92&0.56&0.74\\
\hline  \hline
PSI-based model\cite{ChoiNatComm}&0.144&0.122&13.4&8.4&0.96&1.82&0.69&0.96\\
NeuralNEB\cite{NeuralNEB}&0.136&0.096&6.5&2.1&1.26&1.66&0.77&0.83\\
\bottomrule
\end{tabular}}
\label{table:EvsRMSD}
\end{table*}
\section*{Discussion}
Elucidating TS structures is essential for uncovering the underlying microscopic mechanisms of chemical reactions and estimating reaction barriers for building large reaction networks. 
In this work, we extended SE(3) equivariant diffusion models to respect the object-wise symmetries, leading to OA-ReactDiff, an object-aware SE(3) equivariant diffusion model that first fulfills all the symmetries and constraints for generating elementary reactions.
In addition, we built a confidence-model-enabled recommender to overcome the stochastic nature of diffusion model to select from sampled generated by OA-ReactDiff in multiple runs.
OA-ReactDiff + recommender gives an RMSD of 0.129 Å and MAE of 4.4 kcal/mol compared to the true TS structure obtained by computationally demanding climbing image NEB calculations.
By further using the confidence score for uncertainty quantification, we can selectively perform climbing image NEB only for 14\% of elementary reactions that OA-ReactDiff is most uncertain about, leading to a reduced MAE of 2.6 kcal/mol.

\qquad The current OA-ReactDiff approach has two major limitations.
First, we describe an elementary reaction as a set of 3D structures (say $N$ atoms for reactant, TS, and product), which leads to a system that is 3x larger (that is, $3N$ atoms).
Although the most expensive equivariant update is still object-wise (that is, scales with $N$), the scalar message-passing update requires building a fully-connected graph for the $3N$ atoms, which will be the bottleneck for applying OA-ReactDiff on chemical systems > 100 atoms on a single GPU.
Second, despite the workaround of using a confidence model and recommender to select a unique sample generated by OA-ReactDiff, the stochastic nature of diffusion model cannot be avoided.
This leads to uncertainty for the sample quality of generated TS structure and accumulated runtime for running OA-ReactDiff repeatedly.
These limitations are inherent for diffusion models, which can potentially be addressed by reformulating elementary reaction generation as a transport problem, where optimal transport \textit{via} flow matching\cite{FlowMatching} or Schrödinger bridge\cite{I2SB} can be applied.
The authors are actively exploring along this direction as a future work.
\textcolor{black}{
During the revision of this manuscript, we found a concurrent work from Kim et al. named TSDiff \cite{2DTSDiff}, which showcases the use of diffusion model for TS generation in another perspective.
There, the TS structure is generated from a 2D graph composed of molecular connectivity of reactant and product via a diffusion model. 
They circumvent the need to handle object-wise SE(3) symmetry by removing 3D conformations of reactant and product as inputs. 
This simplification comes with the price of not being able to select “the best” TS structure out of all generated samples \cite{QiyuanNCS2021}.
However, it would be of interest to compare OA-ReactDiff plus 2D-to-3D conformation sampling and TSDiff in reaction exploration in future work. 
}

\qquad Together with uncertainty quantification, OA-ReactDiff + recommender reached both the structural and energetic accuracy required in TS search, which can be readily integrated in current high throughput computation workflows for reaction network exploration. 
In this work, we focus on the relatively well defined TS search problem such that we can evaluate our newly-developed OA-ReactDiff more easily and demonstrate the promise  of this new model. 
OA-ReactDiff, however, models the joint distribution of structures in elementary reactions and thus is not limited to double-ended TS search problem and can be applied in single-ended (i.e, only the reactant is provided) or zero-ended (that is, only the chemical composition of a system is provided) scenarios.
\textcolor{black}{
Although we perform DFT calculations for evaluating the barrier height throughout this work, OA-ReactDiff framework can be readily adopted to predict the reaction barrier by swapping the model output layer.
}
Very recently, a more diverse elementary reaction dataset 17 times larger than Transition1x, named as RGD1, has been established \cite{RGD1}.
Provided that the quality of diffusion model is highly dependent on the size of training data, RGD1 has the potential of unleashing the power of OA-ReactDiff for establishing large reaction networks and exploring chemical reactions with unknown mechanisms with a greatly reduced number of DFT calculations.
Lastly, despite solely focusing on chemical reactions, the object-aware SE(3) equivariant diffusion model developed in this work can be applied to diverse chemical problems where the system of interest consists of multiple 3D objects, in which their interactions do not depend on their locations in Euclidean space.

\section*{Methods}

\paragraph{Equivariant diffusion models.}\label{edm}
\textit{Equivariance.-- } A function $f$ is said to be equivariant to a group of actions $G$ if $g\circ f(x) = f(g\circ x)$ for any $g\in G$ acting on $x$~\cite{equivariance, gdlbook}. 
In this paper, we specifically consider the Special Euclidean group in 3D space (SE(3)) which includes permutation, translation and rotation transformations. 
We intentionally break the reflection symmetry so that our model can describe molecules with chirality. 

\textit{Diffusion models.-- } Diffusion models are originally inspired from non-equilibrium thermodynamics~\cite{ddpm, diffusion2015,scoresde}. 
A diffusion model has two processes, the forward (diffusing) process and the reverse (denoising) process. 
The noise process gradually adds noise into the data until it becomes a prior (Gaussian) distribution:
$$
q(x_t|x_{t-1}) = \mathcal{N}(x_t|\alpha_t x_{t-1}, \sigma_t^2 I),
$$
where $\alpha_t$ controls the signal retained and $\sigma_t$ controls the noise added. 
A signal-to-noise ratio is defined as $\text{SNR}(t)=\frac{\alpha_t^2}{\sigma_t^2}$. We set $\alpha_t = \sqrt{1-\sigma_t^2}$ following the variance preserving process in~\cite{scoresde}. 

\qquad The \textit{true denoising process} can be written in a closed form due to the property of Gaussian noise:
\begin{align}
q(x_s|x_0, x_{t}) &= \mathcal{N}(x_s|\mu_{t\rightarrow s}(x_0, x_t), \sigma^2_{t\rightarrow s}I), \nonumber\\
\mu_{t\rightarrow s}(x_0, x_t) &= \frac{\alpha_{t|s}\sigma^2_s}{\sigma^2_t}x_t + \frac{\alpha_s\sigma^2_{t|s}}{\sigma^2_t}x \;\; \text{and} \;\; \sigma_{t\rightarrow s} = \frac{\sigma_{t|s}\sigma_{s}}{\sigma_t}, \nonumber
\end{align}
where 
\textcolor{black}{
$s$ < $t$ refer to two different timesteps along the diffusion/denoising process ranging from 0 to T,
}
$\alpha_{t|s} = \frac{\alpha_{t}}{\alpha_{s}}$, $\sigma^2_{t|s} = \sigma^2_t - \alpha^2_{t|s} \sigma^2_{s}$.
However, this \textit{true denoising process} is dependent on $x_0$ which is the data distribution and not accessible.
Therefore, diffusion learns the denoising process by replacing $x_0$ with $\hat{x} = \epsilon_{\theta}(x_t, t)$ predicted by a denoising network $\epsilon_{\theta}$.
The training objective is to maximize the variational lower bound (VLB) on the likelihood of the training data:
$$
-\log p(x) \leq D_{KL}(q(x_T|x_0)||p_{\theta}(x_T)) - \log p(x_0|x_1) + \Sigma_{t=2}^T D_{KL}(q(x_{t-1}|x_0, x_t)|| p_{\theta}(x_{t-1}|x_t))
$$
Empirically, a simplified objective has been found to be efficient to optimize~\cite{ddpm}:
$$
\mathcal{L}_\mathrm{simple} = \frac{1}{2}||\epsilon - \epsilon_{\theta}(x_t, t)||^2,
$$

\textit{Equivariant diffusion models.-- } To build an SE(3)-equivariant diffusion model, it has been proven that we need an SE(3)-invariant prior and an SE(3)-equivariant transition kernel~\cite{frank}.
To guarantee equivariance on permutation, rotation, and translation, a necessary condition is to use an SE(3)-equivariant transition kernel (that is denoising network), as we will explain in details at a later section. (see \textit{~\nameref{leftnet}}). 
There are additional requirements for rotation and translation.
For rotations, the isotropic Gaussian prior has the nice property to transform equivariantly.
For translations, we need to limit the distribution on the linear subspace where the center of mass is the origin~\cite{frank}. 

\paragraph{Inpainting for conditional generation.}\label{inpainting}
Inpainting is a flexible technique to formulate the conditional generation problem for diffusion models. \cite{Repaint}
Instead of modeling the conditional distribution, inpainting models the joint distribution during training.
During inference, inpainting methods combine the conditional input as part of the context through the noising process of the diffusion model before denoising both the conditional input and the inpainting region together. The resampling technique~\cite{Repaint} has demonstrated excellent empirical performance in harmonizing the context of the denoising process as there is sometimes mismatch between the noised conditional input and the denoised inpainting region. Specifically, resampling increases the total number of sampling steps in each denoising step by sampling the inpainting region back and forth together with the conditional input.
Despite resampling increases the number of total denoising steps, this can be compensated by decreasing the number of total denoising steps accordingly by striding the sampling schedule~\cite{iddpm} without significantly sacrificing the model performance.

\paragraph{LEFTNet.}\label{leftnet}
We build our denoising network on top of a recently proposed SE(3)-equivariant GNN, LEFTNet~\cite{leftnet}. 
The main idea of LEFTNet relies on building local frames to scalarize the vector (for example position, velocity) and higher order tensor (for example stress) which becomes invariant to SE(3) transformations. 
Tensorization can be applied to invert the scalar back to vector and higher order tensor without information loss in each layer to update these quantities. 
The benefit of scalarization is demonstrated by the flexibility of neural network parameterizations without breaking the symmetry and further proved by the universal approximation theorem~\cite{clofnet} such that the resulting neural network has the universality in the space of continuous SE(3) and permutation equivariant functions. 

\textit{Scalarization and tensorization.-- } Scalarization and tensorization are two operations in differential geometry to convert geometric quantities. 
Specifically, scalarization transforms geometric tensors into scalars while tensorization is the inverse of scalarization, transforming scalars back to geometric tensors. 
In this case, scalarization is used to transform equivariant quantities by three equivariant orthonormal frames:
$$
\mathcal{F}:= (e_1, e_2, e_3)
$$
For simplicity, we use vector as an example. The geometric tensors are scalarized by the inner product between the frames $\mathcal{F}$ and the input vector $x$ as follows:
$$
\mathbf{x}:=(x\cdot e_1, x\cdot e_2, x\cdot e_3)
$$
On the contrary, tensorization reverses the process by:
$$
x:= \mathbf {x_1} e_1 + \mathbf {x_2} e_2 + \mathbf {x_3} e_3.
$$
where $(\mathbf {x_1}, \mathbf {x_2}, \mathbf {x_3})$ is the input scalar tuple and $x$ is the converted tensor.

\textit{Message passing neural network (MPNN).-- } A molecular graph can be represented as $\mathcal{G}=(\mathcal{V}, \mathcal{E})$ where $\mathcal{V}$ is a set of nodes or atoms and $\mathcal{E} \subseteq \mathcal{V}\times \mathcal{V}$ is a set of edges or bonds connecting pairs of nodes.
For each node, we have atom types $h \in \mathbb{R}^{n\times k}$, where $n$ denotes number of nodes and $k$ denotes number of atom types. Edge features are attached to the edges connecting nodes $i$ and $j$ as $e_{ij}$. 
Message passing neural network is a common framework to learn embeddings over graphs~\cite{mpnn}. 
MPNNs often have three parts: (1) message, (2) update and (3) readout. 
The common message between each pair of nodes are:
$$
m_i = \sum_{j\in \mathcal{N}(i)} M(h_i, h_j, e_{ij}),
$$
where $M$ is the message function, $\mathcal{N}(i)$ denotes neighbors of node $i$ and $m_i$ is the message.
Then the message is used to update the node feature as:
$$
h_i = U(h_i, m_i),
$$
where $U$ is the update function. After a number of layers, the global embedding is calculated by:
$$
g = R(h_i|i\in \mathcal{G}),
$$
where $R$ is the readout function. 

\qquad In our case, the molecular graph also has atomic coordinates $x \in \mathbb{R}^{n\times 3}$ and the message function of the MPNN needs to be equivariant to SE(3) transformations. It is achieved by moving the center of mass to the origin (translations) and incorporating the scalarized coordinates $s_{ij}$ in the message function (rotations):
$$
m_i = \sum_{j\in \mathcal{N}(i)} M(h_i, h_j, s_{ij}, e_{ij}),
$$
where $s_{ij}$ is obtained by scalarizing the input coordinate over the frames $\mathcal{F}_{ij}$ between each pair of nodes $i$ and $j$:
\begin{align}
e_1 = \frac{x_i - x_j}{||x_i - x_j||},& e_2 = \frac{x_i \times x_j}{||x_i \times x_j||}, e_3 = e_1 \times e_2, \nonumber\\
s_{ij} &= (x_i \cdot e_1, x_i \cdot e_2, x_i \cdot e_3)\nonumber
\end{align}

\textit{Building towards LEFTNet.-- } Motivated by distinguishing local 3D geometric isomorphisms, LEFTNet introduced a local structure encoding module to encode the local atomic environment of each atom in the scalarization operation. In addition, LEFTNet designed another frame transition encoding block to consider the transition between two frames (the atom and its neighbor atom) when calculating the message between them. 

\paragraph{Object-aware SE(3) implementation.}\label{oa_details}
In general, a system consists of multiple objects (molecules or proteins) that do not have interactions through the 3D Euclidean space can be described by a set of independent graphs, $\{\mathcal{G}_i = (\mathcal{V}_i, \mathcal{E}_i)\}$.
In elementary reaction, for example, we have three objects which can index, reactant ($i=0$), TS ($i=1$), and product ($i=2$).
One important addition symmetry for these systems is object-wise SE(3) equivariance, meaning any SE(3) transformation on each individual object in a system should not influence its description:
$$
\{g_i\circ f(\mathcal{G}_i)\} = \{f(g_i\circ \mathcal{G}_i)\}
$$
$g_i$ represents an SE(3) transformation on $i^{th}$ object, which is not necessarily the same for all objects. 
On the other hand, any non-SE(3) transition on any objects in a system should influence its description:
$$
\{g_0\circ f(\mathcal{G}_0), ..., q_i\circ f(\mathcal{G}_i), ..., g_n\circ f(\mathcal{G}_n)\} \neq \{f(g_0\circ \mathcal{G}_0), ..., f(q_i\circ \mathcal{G}_i), ..., f(g_n\circ \mathcal{G}_n)\}
$$
where $q_i$ represents a non-SE(3) transformation on $i^{th}$ object.
An SE(3) GNN would hold for the latter but violate the former symmetry.

\qquad To simultaneously fulfill these two requirements (along with other symmetries naturally fulfilled by SE(3) equivariant diffusion models), we developed a generic approach to adapt any SE(3) GNN to its object-aware SE(3) equivalent.
The essence of this approach is to still perform equivariant update using an SE(3) GNN for individual object, but only allow scalar-type message passing among different objects to avoid the "leak" of their relative position information.
Starting from the original graph representations, $\{\mathcal{G}_i = (\mathcal{V}_i, \mathcal{E}_i)\}$, $\mathcal{G}_i$ first gets updated by an SE(3) equivariant block for message passing. 
The resulting graphs ($\mathcal{G}_i^\prime$) of all objects are scalarized and concatenated as a system-level fully-connected graph with only scalar representation $\Tilde{\mathcal{G}}$.
All the scalar node features in $\Tilde{\mathcal{G}}$ is then updated by a scalar message-passing block.
This way, interactions among different objects ($i$) in the system are included without introducing the positioning of different objects as all high-order tensors have been scalarized.
Finally, the updated nodes scalars are combined with the outputs from equivariant update block.
This constitutes an object-aware SE(3) interaction block built on top of a vanilla SE(3) update function.
Similar to SE(3) GNNs, this interaction block repeats several times until the final graph representations are readout.

\paragraph{Details for model training.}\label{training}
\textit{Dataset and train/test partitioning.-- }Built on top a large chemically diverse dataset by Grambow \textit{et al.}\cite{Grambow2020}, Transition1x\cite{ts1x} dataset consists of 10,073 elementary reactions optimized by climbing image NEB.
We partitioned Transition1x randomly, with 9,000 reactions used in training and validation and the remaining 1,073 reactions as set-aside test set.
It is not guaranteed that all species in test reactions are unseen by a trained model due to the overlapping structures in different reactions.
However, due to the uniqueness of elementary reaction, there is, at most, one chemical species (specifically, reactant or product) that may overlap in multiple reactions.
We think this partition is reasonable because all TS structures in the test set are completely new and unseen from model training.
In addition, having a certain degree of overlap in reactants/products for the training and set-aside test set is useful to judge whether the diffusion model only memorizes training samples rather than learning to generate new samples.
\textcolor{black}{
Throughout this manuscript, we do not generate new reaction data through DFT-based TS optimizations, except for ones used as examples in Fig. \ref{fig:example_generation}.
Therefore, all DFT results discussed here are computed with $\omega$B97x/6-31G(d).
}

\textit{OA-ReactDiff training.-- } We trained OA-ReactDiff with LEFTNet as our vanilla SE(3) equivariant GNN.
We used a set of hyperparameters similar to that for QM9 dataset in the original paper\cite{leftnet}, with 96 radial basis functions, 196 hidden channels for message passing, and 6 equivariant update blocks. 
A large neighbor cutoff threshold of 10 Å is used to impose fully connected graphs within each molecule.
We mostly adopted hyperparameters of the diffusion process from the EDM paper\cite{EDM}, where a \textcolor{black}{second-order} polynomial noise schedule \textcolor{black}{(that is, $\alpha_t = 1 - (t / T)^2$)} and $L_{\mathrm{simple}}$ loss function is used.
We observed a marginal improvement in model performance as we increase the total diffusion steps and used 5000 steps for our final model.
We used a learning rate of 0.0005 and a batch size of 32, which is the largest batch size that we can afford with a V100/16GB GPU.
The OA-ReactDiff model was trained for 2,000 epochs. 
During the training, the 9,000 reactions were further partitioned by a 8:1 ratio as training and validation.
In practice, however, we observed that early stopping is not required due to the near monotonic decreasing loss for both the training and validation data during the entire training process.

\textit{Confidence model training.-- } The confidence model shares exactly the same set of hyperparameters as the scoring network, with the only change being the use of \textit{sigmoid} function at the final output layer.
To get data for training the confidence model, we ran OA-ReactDiff on the 9,000 training reactions for 40 runs, generating 360,000 synthetic reactions.
We labeled a reaction as "good" (that is, 1) if the generated TS structure has a RMSD < 0.2 Å compared to the true TS, and labeled it as "bad" (that is, 0) otherwise.
Lastly, we train the confidence model as a binary classifier, where the predicted probability is used as the confidence score to estimate the quality of generated TS structure.
Note that we used the same partition for both scoring network and confidence model, which ensures the 1,073 reactions in the set-aside test set are unseen to both models during evaluation.
\section*{Code availability}
Code for OA-ReactDiff is available as a open source repository on github, https://github.com/chenruduan/OAReactDiff

\section*{Acknowledgements}
This work was supported by the U.S. Office of Naval Research under grant no. N00014-20-1-2150 (C.D. and H.J.K.) and National Science Foundation grant CBET-1846426 (H.J. and H.J.K.). 
C.D. thanks the Molecular Sciences Software Institute for the fellowship support under NSF grant OAC-1547580. 
C.D. thanks Q. Zhao and M. Monkey for discussions about elementary reactions. 
C.D. thanks A. Nandy and W. Du for discussions about equivariant graph neural networks. 
C.D. and Y.D. thank G.-H. Liu and T. Chen for discussions about diffusion model and Schrödinger bridge. 
C.D. and H.J. thank Y. Zhao for his help on preparing a demo jupyter notebook for this work.
The authors thank S. Choi and M. Schreiner for communications and providing their raw data that makes the comparison in Table 1 possible.

\section*{Author contributions}
C.D.: conceptualization, methodology, software, validation, investigation, data curation, writing of original draft, review and editing, and visualization. 
Y.D.: methodology, software, writing of original draft, and review and editing. 
H.J.: data curation, review and editing. 
H.J.K.: writing of original draft, review and editing.

\section*{Competing interests}
The authors declare no competing financial interest at this moment.

\bibliographystyle{naturemag_doi} 
\bibliography{main.bib}

\clearpage

\setstretch{1}

\appendix

\title{\textit{Supplementary Information} for "Accurate transition state generation with an object-aware equivariant elementary reaction diffusion model"}

\maketitle

\renewcommand{\thesection}{S\arabic{section}}  
\renewcommand{\thetable}{S\arabic{table}}  
\renewcommand{\thefigure}{S\arabic{figure}}
\setcounter{figure}{0}
\setcounter{table}{0}

\makeatletter
\renewcommand{\fnum@figure}{\textbf{Figure \thefigure}. }
\renewcommand{\fnum@table}{\textbf{Table \thetable. }}

\section*{Abbreviation}
The following is the list of abbreviation utilized in the main paper.
\begin{enumerate}
    \item OA-ReactDiff: \underline{O}bject-\underline{a}ware SE(3) GNN for generating sets of 3D molecules in elementary \underline{react}ions under the \underline{diff}usion model
    \item RMSD: Root mean square deviation.
    \item SE(3): Special Euclidean group in 3D space.
    \item TS: Transition state.
    \item MAE: Mean absolute error.

\end{enumerate}

\section{Physical symmetries and constraints in an elementary reaction.}
\label{SI:required_symmetries}
An elementary reaction that consists of $n$ fragments as reactant and $m$ fragments as product can be described as $\{\mathrm{R}^{(1)}, ..., \mathrm{R}^{(n)}, \mathrm{TS}, \mathrm{P^{(1)}}, ...,  \mathrm{P^{(m)}}\}$. This reaction requires the following symmetries:
\begin{enumerate}
    \item \textit{Permutation symmetry among atoms in a fragment}. For any fragment in $\mathrm{R}^{(i)}, \mathrm{TS}, \mathrm{P^{(j)}}$, change of atom ordering preserves the reaction.
    \item \textit{Permutation symmetry among fragments in reactant and product}. The change of ordering in $\{\mathrm{R}^{(1)}, ..., \mathrm{R}^{(n)} \}$ and $\{\mathrm{P}^{(1)}, ..., \mathrm{P}^{(m)} \}$ preserve the reaction.
    \item \textit{Rotation and translation symmetry for each fragment}. Rotation and translation operations on any fragment (i.e., $\mathrm{R}^{(i)}, \mathrm{TS}, \mathrm{P^{(j)}}$) preserve the reaction.
    
\end{enumerate}

\begin{table*}[th]
\centering 
\caption{\textbf{Ablation studies comparing OA-ReactDiff performance on RMSD evaluation with different models}.
Vanilla SE(3) LEFTNet\cite{leftnet} is shown to demonstrate the importance of preserve object-wise symmetry in elementary reaction. EGNN\cite{EGNN} is shown to reflect the importance of vanilla SE(3) model.
}
\resizebox{0.4\textwidth}{!}{
\begin{tabular}{l|cc}\toprule
\multicolumn{1}{c|}{Approach} &\multicolumn{2}{c}{RMSD (Å)} \\\midrule
&mean &median\\\midrule
Object-aware SE(3) LEFTNet&0.183&0.076\\
Vanilla SE(3) LEFTNet&0.638&0.620\\
Object-aware SE(3) EGNN&0.372&0.360\\
\bottomrule
\end{tabular}}
\label{SI:ablation}
\end{table*}

\begin{figure*}[t!]
    \includegraphics[width=1.0\textwidth]{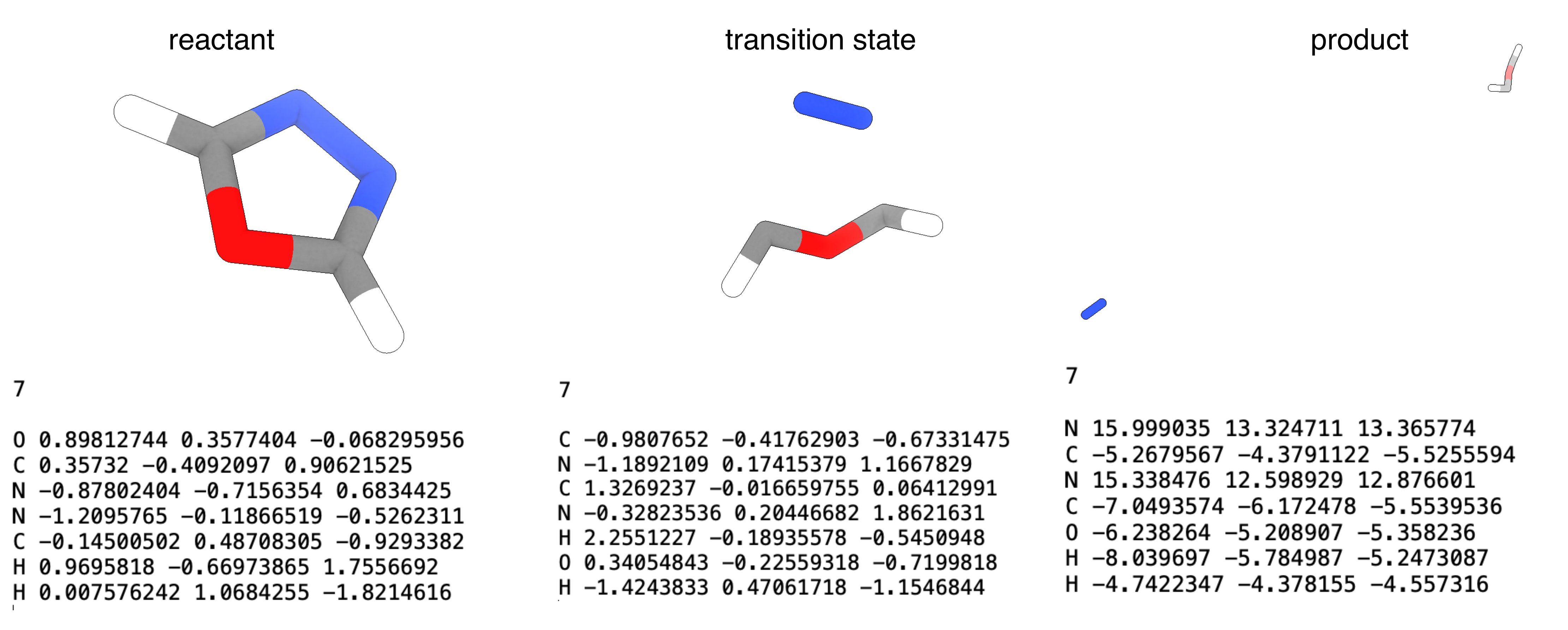}
    \caption{\textbf{Mutil-molecular elementary reactions sampled from OA-ReactDiff by specifying reactant and product.} Here, the atom mapping and fragment alignment are randomized to demonstrate the capability of OA-ReactDiff not relying on these factors. The generated TS structure only has a RMSD of 0.03 Å compared to the DFT ($\omega$B97x/6-31G(d)) optimized true TS. Atoms are colored as follows: gray for C, blue for N, red for O, and white for H.
    }
    \label{Supp:example_multi_molecule_rxn}
\end{figure*}

\begin{figure*}[t!]
    \includegraphics[width=0.9\textwidth]{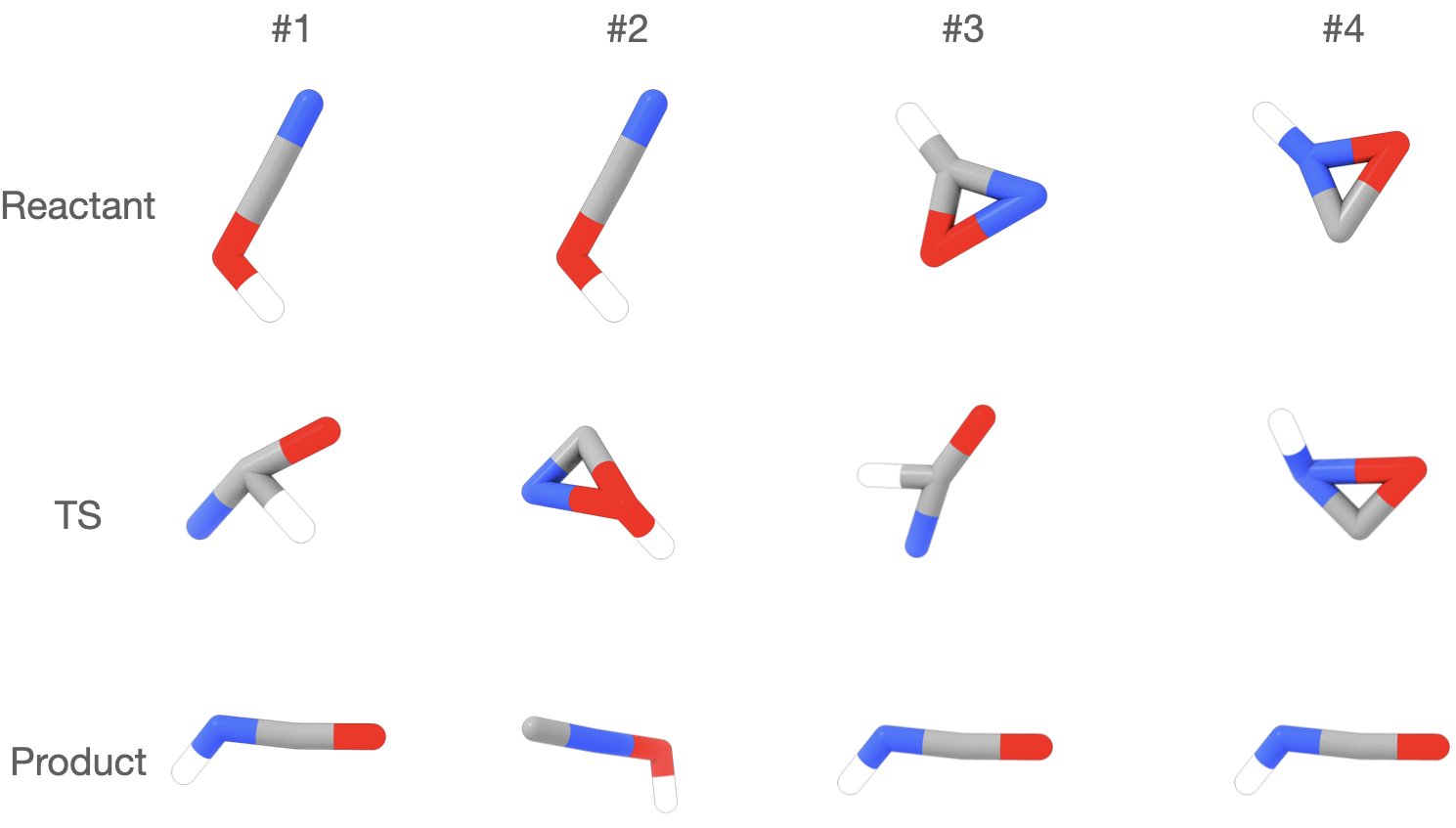}
    \caption{\textbf{Elementary reactions sampled from OA-ReactDiff by only specifying the chemical composition of interest.} Here, we consider a system that contains one C, H, N, and O is chosen. This chemical composition is absent in the Transition1x dataset, and thus is completely new to the trained OA-ReactDiff model. Atoms are colored as follows: gray for C, blue for N, red for O, and white for H.
    }
    \label{Supp:unconditional_samples}
\end{figure*}

\begin{figure*}[t!]
    \includegraphics[width=0.7\textwidth]{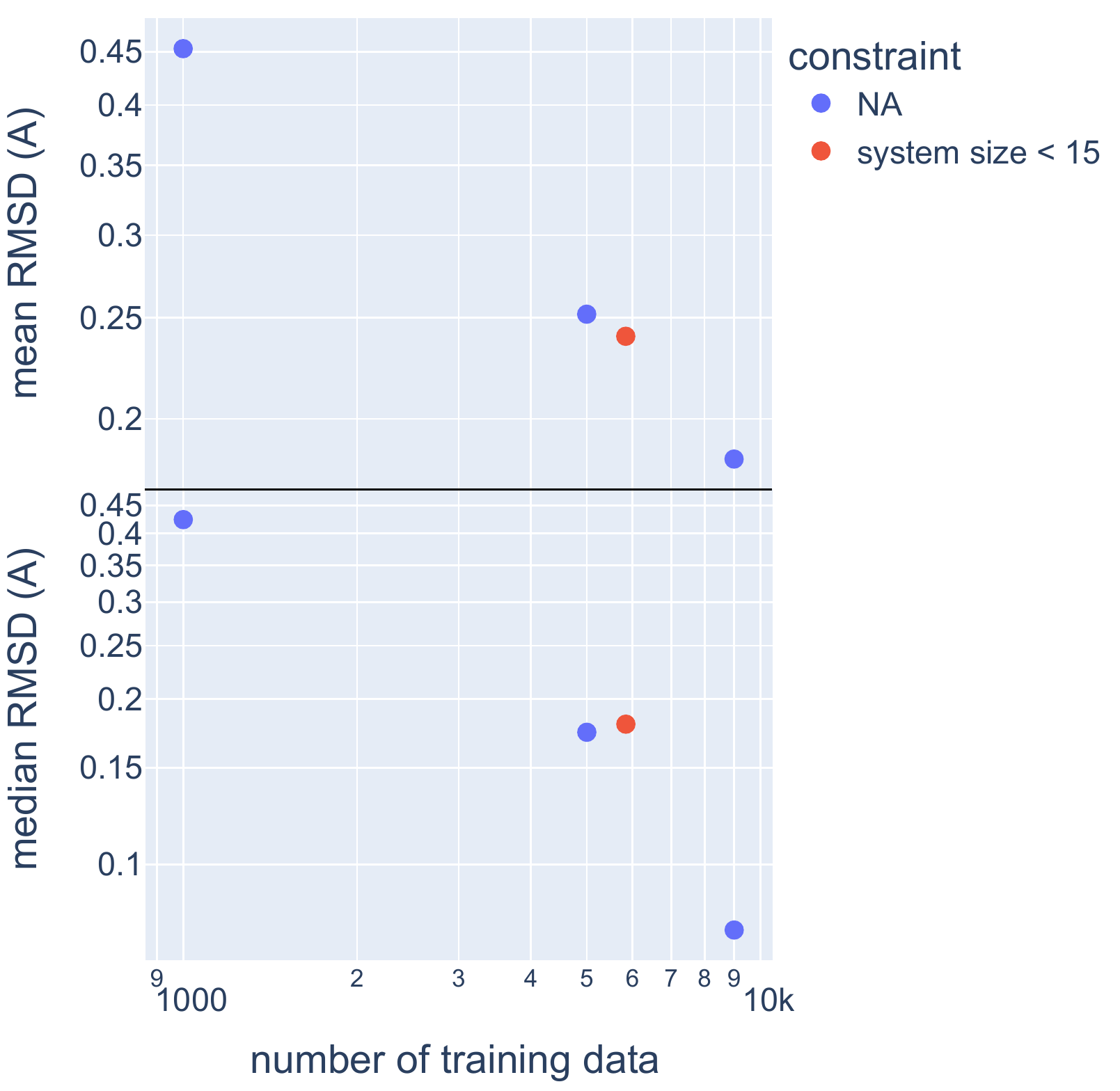}
    \caption{\textbf{RMSD vs. the number of training data}.
    Mean (top) and median (bottom) RMSD for OA-ReactDiff models trained on different number of training data that are either randomly sampled (blue) or sampled under the constraint that only systems within 15 atoms in size (red) from the original 9000 training reactions.
    }
    \label{Supp:rmsd_scaling}
\end{figure*}

\begin{table*}[th]
\centering 
\caption{\textbf{Runtime and GPU memory consumption with different batch sizes}.
}
\resizebox{0.8\textwidth}{!}{
\begin{tabular}{l|ccccccccc}\toprule
batch size&1&2&4&8&16&32&64&128&256\\
walltime (sec)&17.1&18.9&28.0&44.7&87.1&171.8&322.9&582.1&1106.1\\
runtime per sample (sec)&17.1&9.5&7.0&5.6&5.4&5.4&5.1&4.6&4.3\\
GPU memory (GB)&1.1&1.3&1.4&1.6&2.2&3.6&5.7&11.2&21.8\\
\bottomrule
\end{tabular}}
\label{SI:runtime_vs_bs}
\end{table*}

\begin{table*}[th]
\centering 
\caption{\textbf{Performance of OA-ReactDiff at different number of training data and constraints}.
}
\resizebox{0.65\textwidth}{!}{
\begin{tabular}{l|cccc}\toprule
number of sample&1000&5000&5734&9000\\
mean RMSD (Å)&0.453&0.252&0.240&0.183\\
median RMSD (Å)&0.424&0.174&0.180&0.076\\
constraints in training data&--&--&system size < 15 atoms&--\\
\bottomrule
\end{tabular}}
\label{SI:training_data_scaling}
\end{table*}

\begin{figure*}[t!]
    \includegraphics[width=1.0\textwidth]{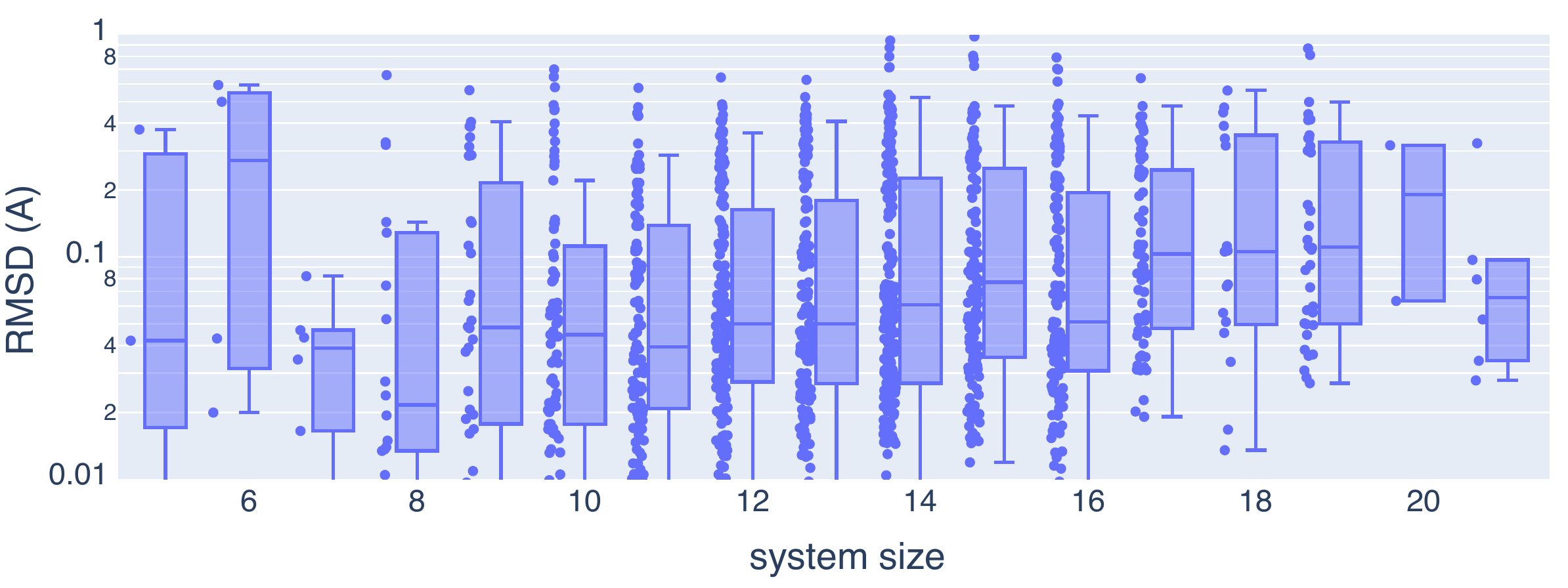}
    \caption{\textbf{Distributions of RMSD binned by system size.}
    The RMSD is computed between top-1 confidence TS structure generated by OA-ReactDiff and the true TS structure for the 1073 test reactions. A standard box plot with median as the horizontal bar is shown with all RMSDs at that system size displayed on its left hand side.
    }
    \label{Supp:rmsd_size_box}
\end{figure*}

\begin{figure*}[t!]
    \includegraphics[width=0.7\textwidth]{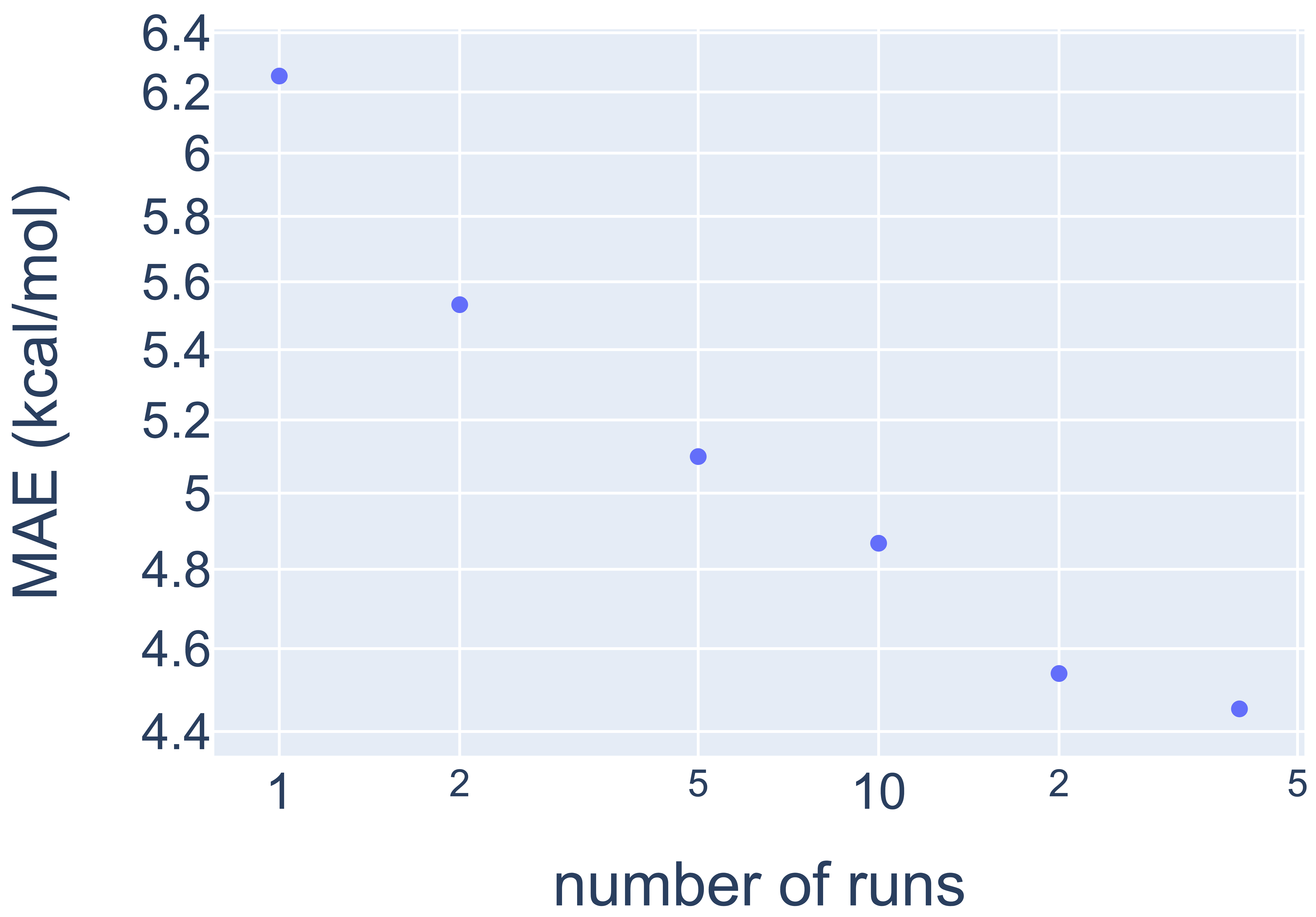}
    \caption{\textbf{Mean absolute energy difference vs. number of runs for OA-ReactDiff sampling.}. A log-log axis is used to shown the near power law dependence. The results are shown on 1073 test elementary reactions.
    }
    \label{Supp:metric_vs_runs}
\end{figure*}

\begin{figure*}[t!]
    \includegraphics[width=0.6\textwidth]{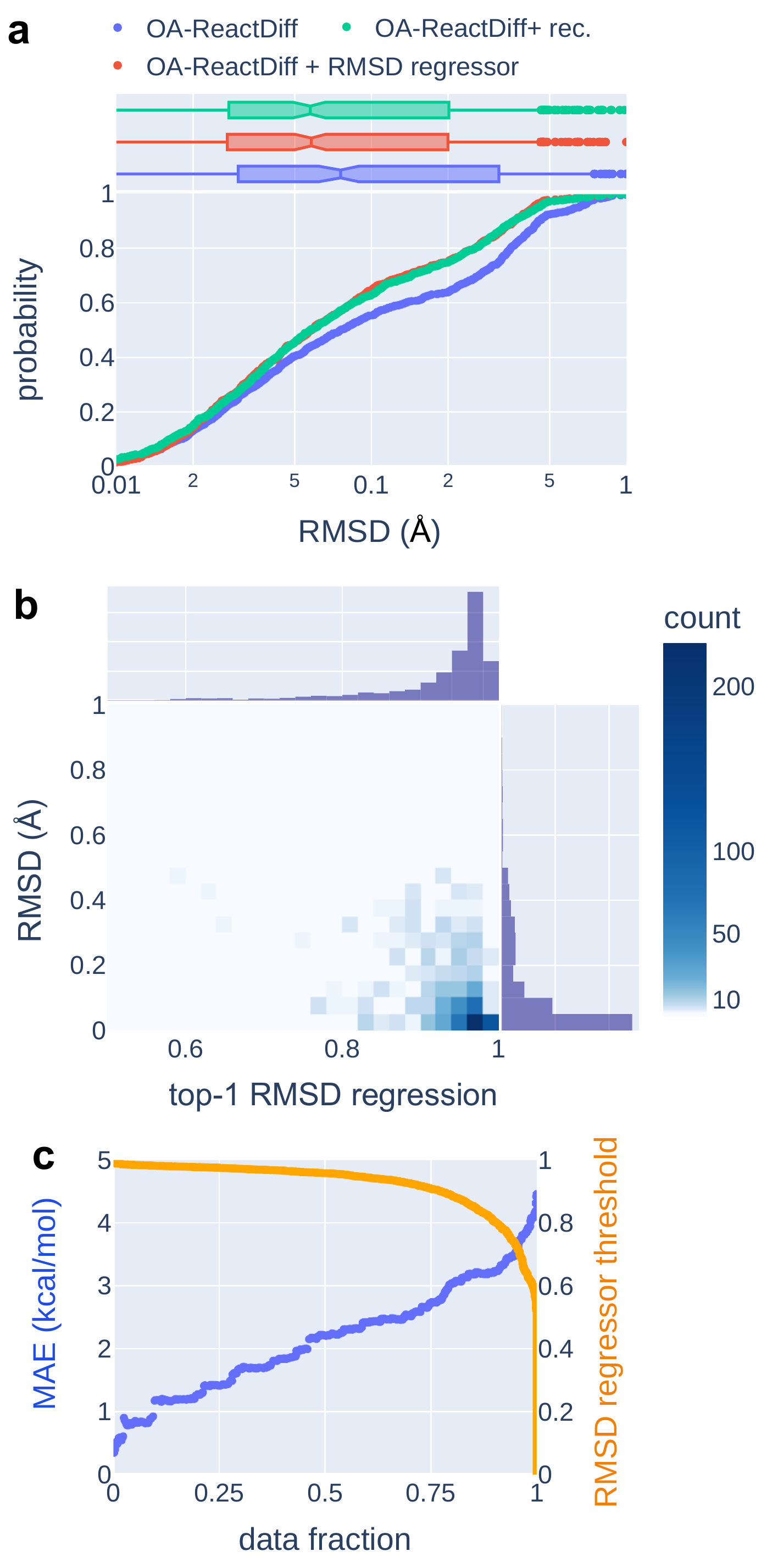}
    \caption{\textbf{Performance of using a RMSD regressor as the confidence model}.
    a. Cumulative probability for RMSD between the true TS structures and OA-ReactDiff samples on
    1073 set-aside test reactions. The OA-ReactDiff samples are evaluated under one-shot generation (blue), the top-1
    confidence sample via classifier-based recommender (green), and the top-1 confidence sample via RMSD regressor out of 40 generated samples for each reaction (red). A log scale of the RMSD is presented for better visibility of the low-RMSD region. 
    b. 2D density map for the RMSD vs. top-1 RMSD regressor confidence for OA-ReactDiff generated samples. A log-scale color gradient is applied to the color bar to reveal low-density areas, which would otherwise be difficult to distinguish. 
    c. MAE of $|\Delta \mathit{E}_\mathrm{TS}|$ (blue, left y-axis) and the corresponding confidence threshold (orange, right y-axis) as a function of the fraction of data considered in the 1073 TS structures selected via the RMSD regressor.
    }
    \label{Supp:regressor_confidence}
\end{figure*}

\begin{figure*}[t!]
    \includegraphics[width=0.7\textwidth]{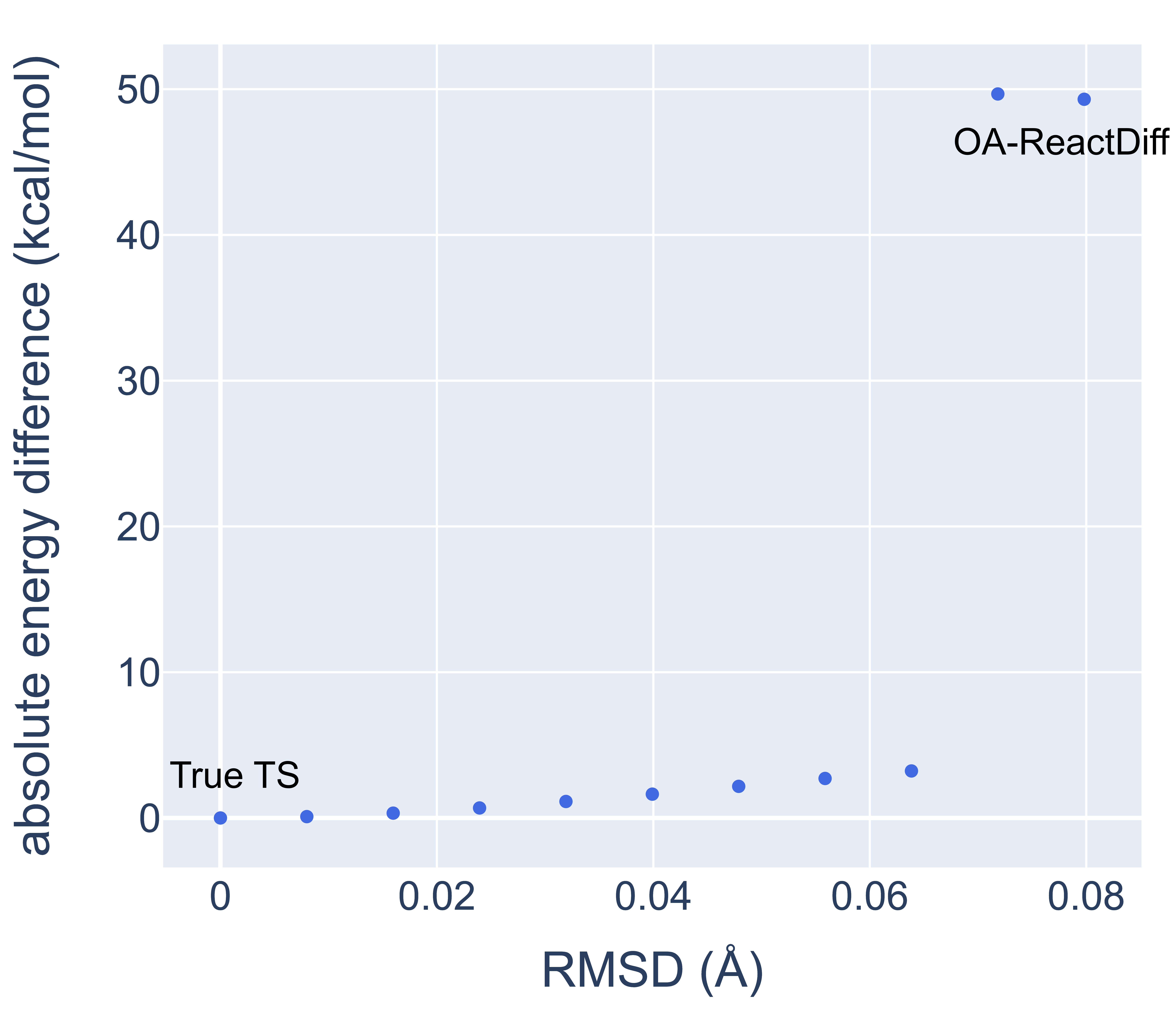}
    \caption{\textbf{Absolute energy difference vs. RMSD for the ten interpolated structure between true (left) and OA-ReactDiff TS (right) for $\mathrm{C_4 H_6 O_2}$.} The abrupt change in energy difference indicates a change in converged electronic state for self-consistent field calculation.
    }
    \label{Supp:C2H6O2}
\end{figure*}

\begin{figure*}[t!]
    \includegraphics[width=0.7\textwidth]{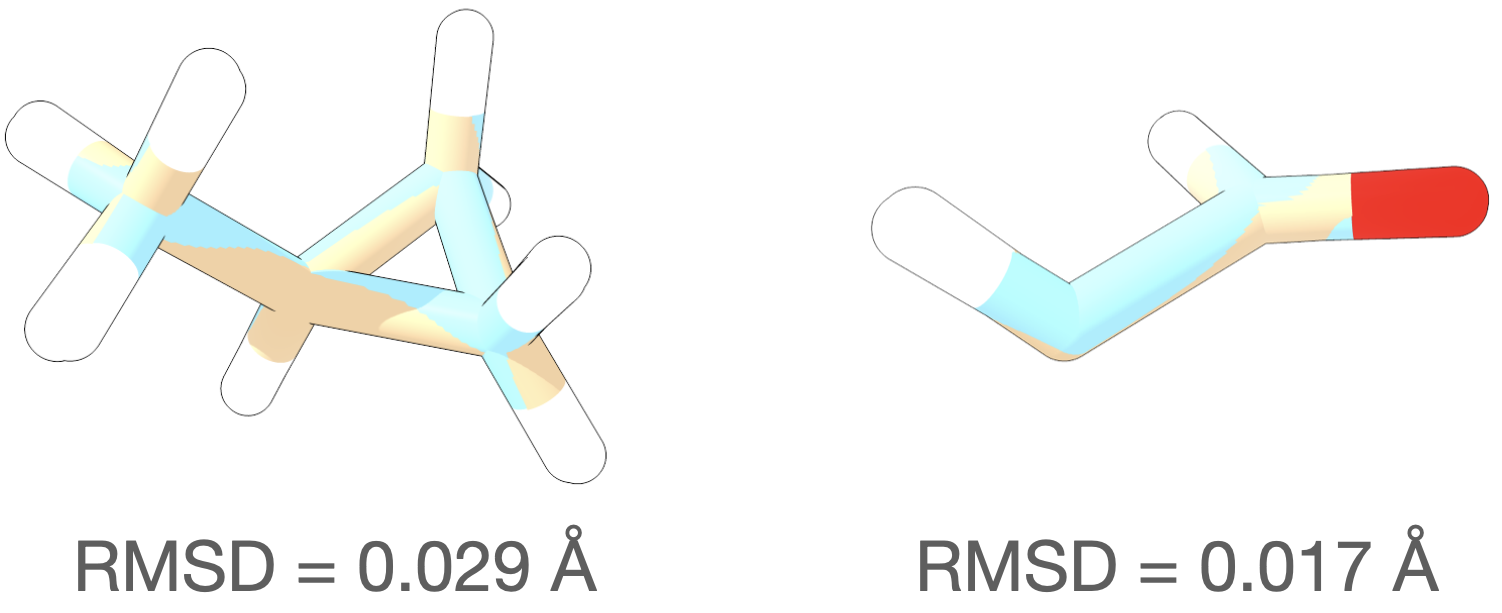}
    \caption{\textbf{Overlapping OA-ReactDiff and true TS structures of $\mathrm{C_6 H_{10} O}$ separated as two fragments and their corresponding RMSD.} Atoms are colored as follows: C in the true TS structure are in tan and those in the OA-ReactDiff sample are in skyblue; O for red, and H for white.
    }
    \label{Supp:C6H10O}
\end{figure*}

\begin{figure*}[t!]
    \includegraphics[width=0.7\textwidth]{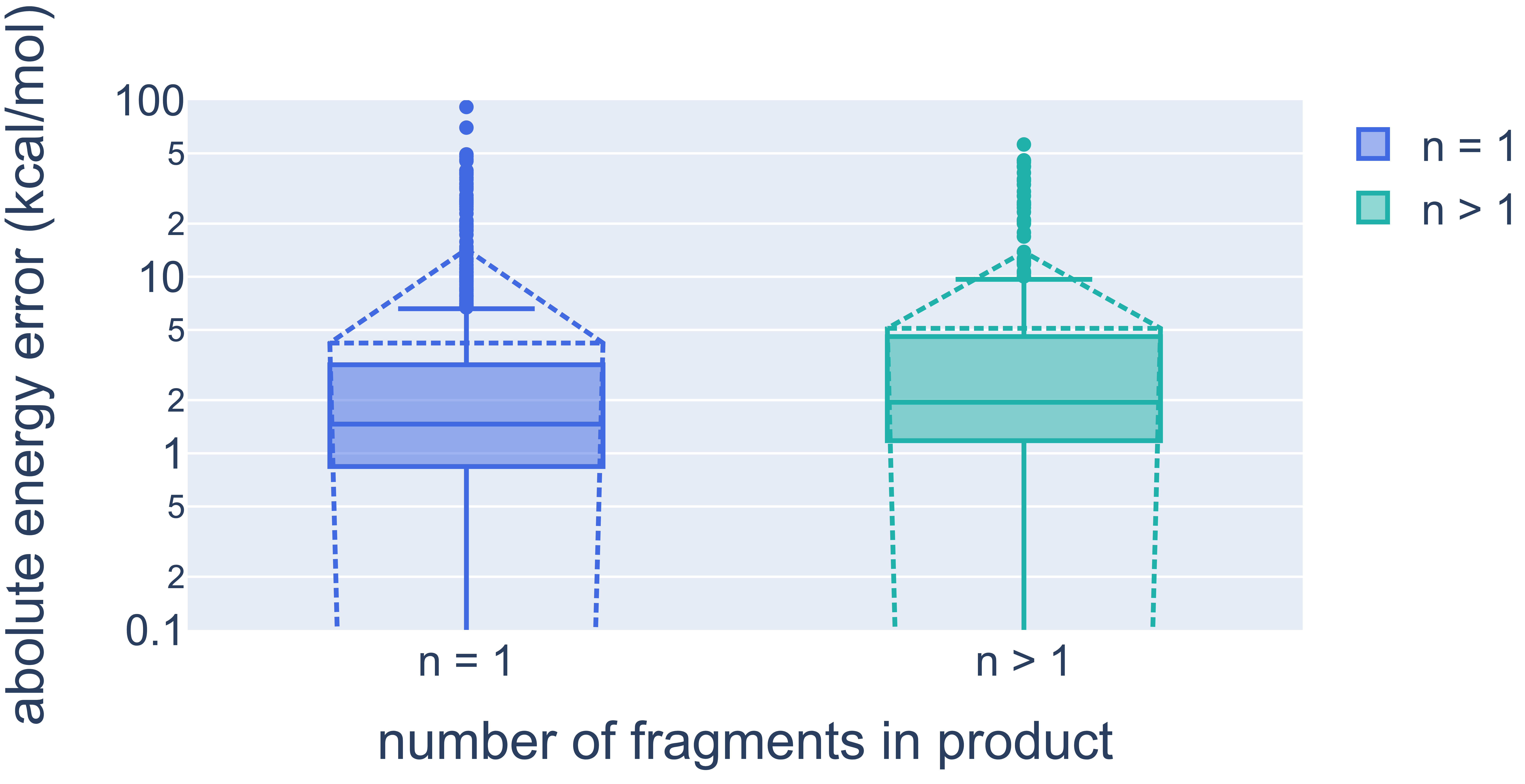}
    \caption{\textbf{Box plot for absolute energy difference of OA-ReactDiff + rec. TS structures grouped by single (i.e., n=1) and multi (i.e., n > 1) product cases.}. The solid lines are shown for the quadrants (Q1, median, and Q3) and the dashed lines are shown for the mean and standard deviation. The results are shown on 1073 test elementary reactions. 
    }
    \label{Supp:num_prod}
\end{figure*}

\begin{table*}[th]
\centering 
\caption{\textbf{Barrier height error of OA-ReactDiff evaluated by different methods}. All energies are reported in the unit of kcal/mol.
}
\resizebox{0.8\textwidth}{!}{
\begin{tabular}{l|ccccccc}\toprule
\multicolumn{1}{c|}{Approach} &\multicolumn{2}{c}{$\omega$B97x}&\multicolumn{2}{c}{B3LYP} &\multicolumn{2}{c}{PBE} &\multicolumn{1}{c}{data fraction}\\\midrule
&mean &median&mean &median&mean &median&\\\midrule
OA-ReactDiff + rec.&4.4&1.6&4.3&1.6&4.9&1.7&1.0\\
OA-ReactDiff + rec. (p>0.5)&3.1&1.4&3.0&1.4&3.7&1.6&0.86\\
\bottomrule
\end{tabular}}
\label{SI:method_dependence}
\end{table*}

\begin{figure*}[t!]
    \includegraphics[width=0.9\textwidth]{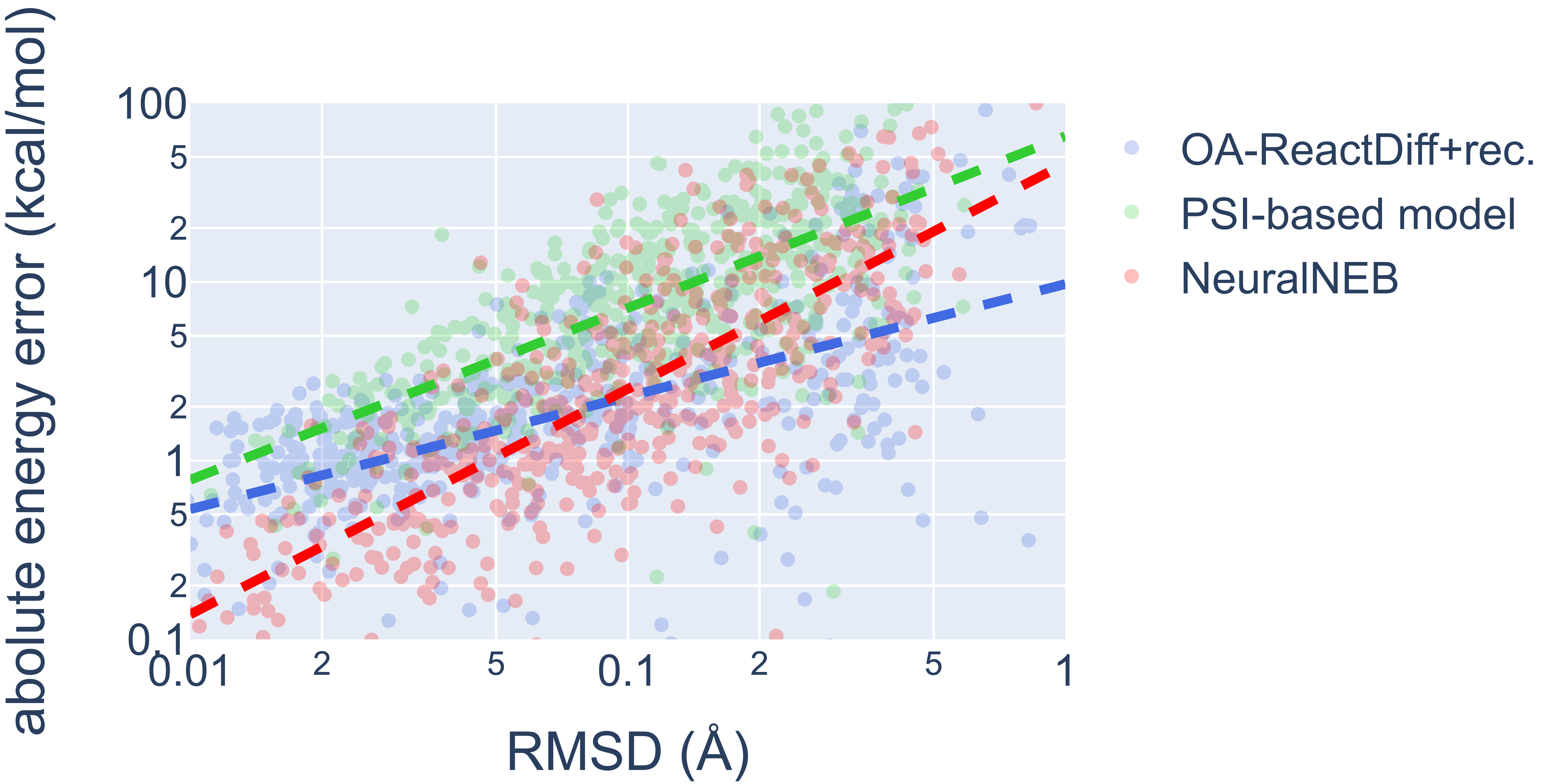}
    \caption{\textbf{Absolute energy difference vs. RMSD and the corresponding linear fit in a log-log plot for OA-ReactDiff + rec, PSI-based model\cite{ChoiNatComm}, and NeuralNEB\cite{NeuralNEB}}.
    }
    \label{Supp:fit_compare}
\end{figure*}

\end{document}